\documentclass[fleqn]{revtex4}

\usepackage{amsmath}
\usepackage{amssymb}
\usepackage{graphicx}
\usepackage{psfrag}

\DeclareMathOperator{\tr}{Tr}

\newcommand{\be}{\begin{equation}} \newcommand{\ee}{\end{equation}}
\newcommand{\ba}{\begin{eqnarray}} \newcommand{\ea}{\end{eqnarray}}
\newcommand{\bea}{\begin{eqnarray}} \newcommand{\eea}{\end{eqnarray}}
\newcommand{\bean}{\begin{eqnarray*}} \newcommand{\eean}{\end{eqnarray*}}
\newcommand{\st}{{\scriptscriptstyle T}}
\newcommand{\sT}{{\scriptscriptstyle T}}

\def\slash#1{\setbox0=\hbox{$#1$}               
        \dimen0=\wd0                            
        \setbox1=\hbox{/} \dimen1=\wd1          
        \ifdim\dimen0>\dimen1                   
        \rlap{\hbox to \dimen0{\hfil/\hfil}}    
        #1                                      
        \else                                   
        \rlap{\hbox to \dimen1{\hfil$#1$\hfil}} 
        /                                       
        \fi}                                    %

\begin{document}

\title{The Construction of Gauge-Links in Arbitrary Hard Processes}

\author{C.J. Bomhof}
\email{cbomhof@nat.vu.nl}
\affiliation{
Department of Physics and Astronomy, Vrije Universiteit Amsterdam,\\
NL-1081 HV Amsterdam, the Netherlands}

\author{P.J. Mulders}
\email{mulders@few.vu.nl}
\affiliation{
Department of Physics and Astronomy, Vrije Universiteit Amsterdam,\\
NL-1081 HV Amsterdam, the Netherlands}
                                                                                
\author{F. Pijlman}
\email{fetze.pijlman@philips.com}
\affiliation{
Department of Physics and Astronomy, Vrije Universiteit Amsterdam,\\
NL-1081 HV Amsterdam, the Netherlands}

\begin{abstract}
Transverse momentum dependent parton distribution and fragmentation 
functions are described by hadronic matrix elements of bilocal products 
of field operators off the light-cone. 
These bilocal products contain gauge-links, 
as required by gauge-invariance. The gauge-links are 
path-ordered exponentials connecting the field operators along a 
certain integration path. 
This integration path is process-dependent,
depending specifically on the short-distance partonic subprocess. 
In this paper we present the technical details needed in the calculation 
of the gauge-links and a calculational scheme is provided to obtain the 
gauge-invariant distribution and fragmentation correlators 
corresponding to a given partonic subprocess.
\end{abstract}
\date{\today}
\maketitle

\section{Introduction}

Gauge-links (also called eikonal or Wilson lines) play an important role in the
description of single spin asymmetries (SSA's).
Since the initial surprise about the large SSA's in hadron-hadron 
collisions~\cite{Adams:1991rw,Adams:1991cs,Bravar:1996ki,
Airapetian:2001eg,Adler:2003pb,Adams:2003fx,Airapetian:2004tw}, 
several mechanisms have been proposed as an
explanation~\cite{Hagiwara:1982cq,Sivers:1989cc,Sivers:1990fh,Qiu:1991pp,
Qiu:1991wg,Collins:1992kk,Kanazawa:2000hz}.
In these mechanisms correlations between spin and intrinsic transverse momentum
of partons play a central role. 
They can be divided into two classes;
the $T$-even and the $T$-odd correlations. 
The latter can give rise to single spin asymmetries. 

A mechanism for generating $T$-odd effects was introduced by
Qiu and Sterman in Ref.~\cite{Qiu:1991pp}. 
They took the contribution of gauge fields at light-cone past/future infinity into account and
found that those matrix elements could lead to single spin asymmetries. 
These matrix elements are referred to as gluonic poles~\cite{Efremov,Kanazawa:2000hz,Ji:1992eu}.
Other mechanisms to generate SSA's had been introduced by Collins~\cite{Collins:1992kk}
and by Sivers~\cite{Sivers:1989cc,Sivers:1990fh}.
The Collins mechanism originates from final state interactions between 
a detected outgoing hadron and its accompanying jet.
These interactions generate $T$-odd fragmentation functions. 
For distribution functions there are no corresponding initial state interactions and,
therefore, 
the Sivers mechanism was expected to vanish.
However, a few years ago Brodsky, Hwang and Schmidt
showed with a model calculation that $T$-odd effects can also be generated in 
the initial state,
reviving the Sivers mechanism~\cite{Brodsky:2002cx}.
At the same time the process dependence of the underlying mechanism was demonstrated~\cite{Collins:2002kn,Brodsky:2002rv}.
The equivalence of those results and the transverse pieces of the
gauge-link in transverse momentum dependent (TMD) correlation functions
were analyzed by Belitsky, Ji, and Yuan in Ref.~\cite{Belitsky:2002sm}.
Generalizing their results it was found by Boer, Mulders and Pijlman in
Ref.~\cite{Boer:2003cm}
that the effects of transverse gauge-links, 
which lead to $T$-odd effects, 
are equivalent to the Qiu-Sterman mechanism, uniting the various mechanisms.

The gauge-links, 
necessary to have gauge-invariant distribution functions, 
have an integration path that turns out to be process-dependent.
The path structure of the gauge-links gives rise to $T$-odd TMD distribution functions.
In~\cite{Bomhof:2004aw} and~\cite{Bacchetta:2005rm} we considered the kind of integration paths that can appear in general scattering processes. 
With an explicit calculation in QED it was shown that a large set of different gauge-links can appear.
Also for QCD several results were given.
In this paper we address some of the technicalities that were skipped
in~\cite{Bomhof:2004aw}.
We also refer the reader to the detailed study of gauge-links by Pijlman~\cite{Pijlman:2006vm}.
In the next section we state the steps through which the gauge-links can be obtained for a general process.
In section~\ref{VOORBEELDEN} some examples are given and in section~\ref{TAKAHASHI} we present the
derivation of the conjectures made in the preceding sections. 
In the appendix we enumerate the gauge-links in processes like proton-proton
scattering, giving the results for $2{\rightarrow}2$ partonic subprocesses. 
In this paper, we will only give a global discussion of the
consequences of our results.
In an earlier paper it was shown how the presence of gauge-links can lead to modified hard parts,
referred to as \emph{gluonic pole cross sections},
in single spin asymmetries in 
$p^\uparrow p{\rightarrow}\pi\pi X$~\cite{Bacchetta:2005rm}.
The results obtained in this paper will enable us to calculate all the gluonic pole cross sections in $p^\uparrow p{\rightarrow}\pi\pi X$ in a forthcoming paper.
Possible issues related to factorization have been addressed in~\cite{Pijlman:2006vm}.

\section{Calculating gauge-links\label{rules}}

For describing hard hadronic processes
we start with the TMD distribution and fragmentation correlators~\cite{Soper:1976jc,Soper:1979fq,Ralston:1979ys,Collins:1981uw,
Collins:1981tt,Collins:1982wa,Jaffe:1991ra,Mulders:1995dh}
\begin{subequations}\label{CorrFunctions}
\begin{align}
\Phi(x,p_\st)
&=\int\frac{\mathrm{d}(\xi{\cdot}P)\,\mathrm{d}^2 \xi_\st}{(2\pi)^3}\ 
e^{ip\cdot\xi}\,\langle P\vert\,\phi^\dagger(0)\,\phi(\xi)\,\vert P\rangle\ ,
\\
\Delta(z,k_\st)
&=\sum_X\frac{1}{4z}\int \frac{\mathrm{d}(\xi{\cdot}P_h)\,
\mathrm{d}^2 \xi_\st}{(2\pi)^3}\ e^{-ik\cdot\xi}\,
\langle 0\vert\,\phi(0)\,\vert P_h;X\rangle\,
\langle P_h;X\vert\,\phi^\dagger(\xi)\,\vert 0\rangle\ ,
\end{align}
\end{subequations}
where the $\phi$ represent color triplet quark fields $\psi_{r\,i}$ 
(Dirac index
$r$ and color index $i$) in quark correlators $\Phi_q$ and $\Delta_q$ 
or color octet gluon field strengths $F^{\mu\nu}_a$ (color index $a$) 
in gluon correlators $\Phi_g$ and $\Delta_g$. The Lorentz indices on
the gluon fields and the Dirac indices on the quark fields will mostly
be suppressed throughout this paper.
For the field strengths we will often use the matrix representation 
$F^{\mu\nu}_{ij} = F^{\mu\nu}_a\,t^a_{ij}$
and similarly for the gauge-fields $A^\mu$.
The $t^a = \frac{1}{2}\lambda^a$ are the color matrices satisfying the commutation relations $[t^a,t^b] = if^{abc}\,t^c$
and normalization $\tr[t^at^b]=T_F\delta_{ab}$.
For the parton momenta we use the Sudakov decompositions with for each 
hadron's momentum a complementary light-like vector $n$ such that
\begin{subequations}
\begin{alignat}{2}
p&=x\,P+\sigma\,n+p_\st\ ,&\qquad\qquad&\text{(incoming particle)}\\
k&=\frac{1}{z}\,P_h+\sigma_h\,n_h+k_\st\ ,&&\text{(outgoing particle)}
\end{alignat}
\end{subequations}
with $x=p{\cdot}n/P{\cdot}n$, $p_\st{\cdot}P=p_\st{\cdot}n = 0$,
$\sigma=(p{\cdot}P{-}x\,M^2)/(P{\cdot}n)$ and similar relations 
for the components of $k$.  

The bilocal products of field operators in the correlation functions 
in~\eqref{CorrFunctions} contain gauge-links 
$\mathcal U^{[C]}(\eta;\xi)$ to render such products properly gauge-invariant.
The gauge-links are path-ordered exponentials
\begin{equation}\label{POE}
\mathcal U^{[C]}(\eta;\xi)
=\mathcal P\exp\Big[\,-ig\int_C\mathrm dz\cdot A(z)\,\Big]\ .
\end{equation}
This gauge-link is a matrix in color space.
The $C$ is an integration path connecting the space-time points $\xi$ and $\eta$.
In TMD correlation functions the requirement of gauge-invariance alone does not uniquely fix this integration path.
The gauge-links are obtained by resumming all diagrams describing the exchange of collinear gluons between the soft and hard parts~\cite{Efremov:1978xm,Boer:1999si}.
Hence, the integration path is fixed by the hard part of the process. 

\begin{figure}
\centering
\begin{minipage}{4cm}
\centering
\hspace{1cm}\includegraphics[width=\textwidth]{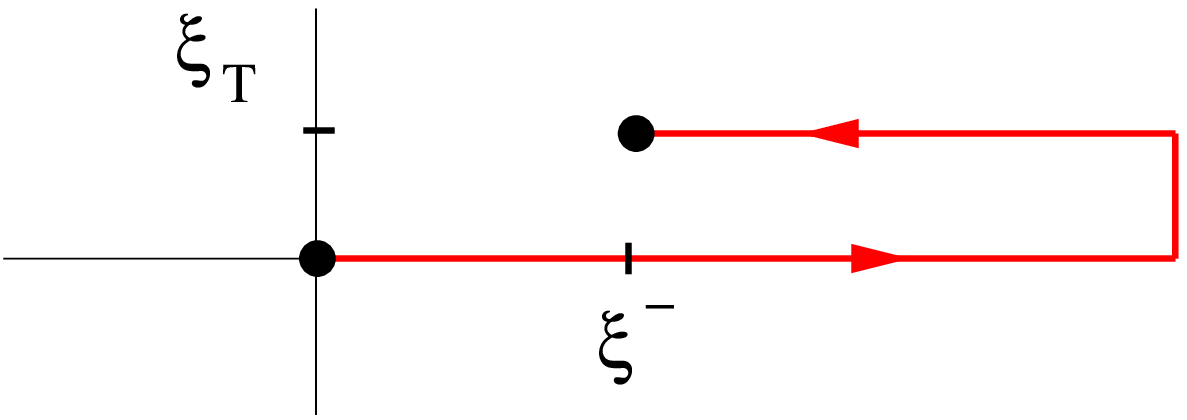}\\[1mm]
(a)
\end{minipage}
\hspace{2cm}
\begin{minipage}{4cm}
\centering
\hspace{1cm}\includegraphics[width=\textwidth]{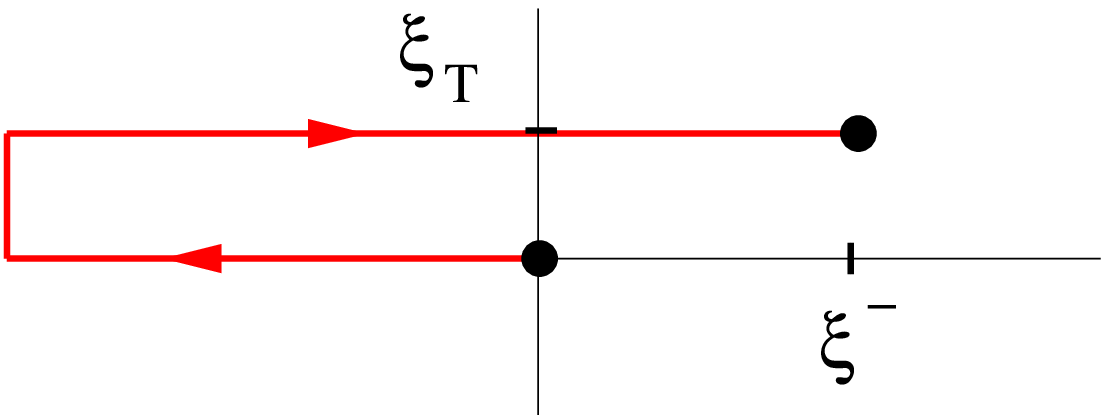}\\[1mm]
(b)
\end{minipage}
\caption{The gauge-link structure in the 
correlator $\Phi$ in (a) SIDIS: $\mathcal U^{[+]}$ and 
(b) DY: $\mathcal U^{[-]}$.\label{simplelinks}}
\end{figure}

All gauge-links that we will encounter are composed of Wilson lines in 
the light-cone and transverse directions:
\begin{align}\label{WilsonLines}
U_{[a;b]}^n
=\mathcal P\exp\Big[-ig\int_a^b\mathrm dz\ n\cdot A(z)\,\Big]\ ,
\qquad\text{and}\qquad
U_{[a;b]}^\sT
=\mathcal P\exp\Big[-ig\int_a^b\mathrm dz_\st\cdot A_\st(z)\,\Big]\ .
\end{align}
The space-time points consist of light-like and transverse components 
$a{=}(a^-,\boldsymbol a_\sT)$ and $a^+{\equiv}a{\cdot}n{=}0$.
Here the vector $n$ is the light-like vector complementary to the momentum of the parent/daughter hadron of the gluon $A$.
Hence, this vector might be different in correlation functions corresponding to different hadrons.
The gauge-links in the distribution functions in semi-inclusive deep inelastic scattering (SIDIS) and 
the Drell-Yan (DY) process are $\mathcal U^{[+]}$ and $\mathcal U^{[-]}$,
respectively, where (see Fig.~\ref{simplelinks})
\begin{align}
\mathcal U^{[\pm]}
&=U_{[(0^-,\boldsymbol0_T);(\pm\infty^-,\boldsymbol0_T)]}^n
U_{[(\pm\infty^-,\boldsymbol0_T);(\pm\infty^-,\boldsymbol\infty_T)]}^\sT
U_{[(\pm\infty^-,\boldsymbol\infty_T);(\pm\infty^-,\boldsymbol\xi_T)]}^\sT
U_{[(\pm\infty^-,\boldsymbol\xi_T);(\xi^-,\boldsymbol\xi_T)]}^n\ .\label{GL}
\end{align}
In processes involving more complicated partonic processes than the ones in SIDIS and DY also more complicated link structures can occur. 
Notably, loops 
$\mathcal U^{[\Box]}=\mathcal U^{[+]}\,\mathcal U^{[-]\dagger}
=\mathcal U^{[-]\dagger}\,\mathcal U^{[+]}$ 
may emerge.

The gauge-links are included in the correlation functions 
in~\eqref{CorrFunctions} by taking for $\phi$ parallel displaced 
fields~\cite{Collins:1989gx}.
For instance, in the quark correlator one encounters
\begin{equation}\label{HALLO}
\phi(\xi) \longrightarrow \Psi(\xi) 
\equiv\mathcal U^{[C]}(\eta;\xi)\,\psi(\xi) 
=\mathcal P\exp\Big[-ig\,\int_C\mathrm ds_\mu\ A^\mu\Big]\,\psi(\xi)\ ,
\end{equation}
or infinitesimally, such that $\eta=\xi{+}\mathrm d\eta$,
\begin{equation}\label{infini1}
\Psi(\xi) 
=(\,1 + ig\,\mathrm d\eta_\mu\;A^\mu_a\,t^a\,)\,\psi(\xi)\ .
\end{equation}
Similarly, the field 
$\phi(0)^\dagger \rightarrow 
\Psi^\dagger(0)\equiv\psi^\dagger(0)\,\mathcal U(0;\eta)$
is a parallel displaced field connecting to the same point $\eta$ 
as the field $\phi(\xi)$. 
Therefore, it is the nonlocal color gauge-invariant
operator combination $\psi^\dagger(0)\,\mathcal U(0,\xi)\psi(\xi)$
involving an integration path connecting $\xi$ and $0$ that will appear in the quark correlator.
In the gluon correlator we will encounter the gauge-link structure
\begin{equation}
\phi(\xi)\longrightarrow\mathbb F^{\,\alpha\beta}(\xi)\equiv 
\mathcal U^{[C']}(\eta;\xi)\,F^{\alpha\beta}(\xi)\,
\mathcal U^{[C'']}(\xi;\eta^\prime)
=\mathcal P\exp\Big[-ig\int_{C'}\mathrm ds_\mu\,A^\mu\,\Big]\,
F^{\alpha\beta}(\xi)\,
\mathcal P\exp\Big[-ig\int_{C''}\mathrm ds_\nu\,A^\nu\,\Big]\ ,
\label{cgi-gluon}
\end{equation}
or infinitesimally, 
with $\eta=\xi{+}\mathrm d\eta$ and $\eta'=\xi{+}\mathrm d\eta'$,
\begin{equation}
\mathbb F^{\,\alpha\beta}(\xi)
=\big(\,1 + ig\,\mathrm d\eta_\mu\,A^\mu_b\,t^b\,\big)\,
F_a^{\alpha\beta}(\xi)\,t^a\,
\big(\,1 - ig\,\mathrm d\eta'_\nu\,A^\nu_c\,t^c\,\big)\ .
\end{equation}
The integration paths $C'$ and $C''$ can be different.  
Hence, in general, the $\mathbb F^{\,\alpha\beta}$ 
is not a standard parallel displaced field strength.  
For a standard displaced field strength operator the integration paths and link endpoints in Eq.~\eqref{cgi-gluon} are identical:
\begin{equation}
\mathbb F^{\,\alpha\beta}(\xi)
=\mathcal U^{[C]}(\eta;\xi)\,
F_a^{\alpha\beta}(\xi)\,t^a\,\mathcal U^{[C]}(\xi;\eta)
\equiv
\big(\,\mathcal U^{[C]}(\eta;\xi)\,\big)_{ab}\,F_b^{\alpha\beta}(\xi)\,t^a\ .
\end{equation}
The last step defines the gauge-link in the 8-dimensional adjoint 
representation
\begin{equation}\label{ADJrep}
\big(\,\mathcal U^{[C]}(\eta;\xi)\,\big)_{ab}
=\frac{1}{T_F}\tr\big[\,t^a\,\mathcal U^{[C]}(\eta;\xi)\,
t^b\,\mathcal U^{[C]\dagger}(\eta;\xi)\,\big]\ .
\end{equation}
Infinitesimally it reads
\begin{equation}\label{infini2}
\big(\,\mathcal U^{[C]}(\eta;\xi)\,\big)_{ab}\,F_b^{\alpha\beta}(\xi)
=\big(\,\delta_{ab} + g\,\mathrm d\eta_\mu\,A^\mu_c\,f^{cab}\,\big)\,
F_b^{\alpha\beta}(\xi)\ .
\end{equation}
In order to properly treat transverse momentum dependent correlators,
one needs the more general 3-dimensional matrix representation 
in~\eqref{cgi-gluon} allowing for different gauge-link structures 
left and right of the field strength tensor. In the gluon correlator 
nonlocal color gauge-invariant combinations such as 
$\tr[\,\mathcal U^{[C_1]}(\xi;0)\,F^{\alpha\beta}(0)\,
\mathcal U^{[C_2]}(0;\xi)\,F^{\gamma\delta}(\xi)\,]$
will appear.

\begin{table}
\begin{tabular}{r|c|c|c|}
&&&contribution to\\
&`free' wave functions&fields in correlator&other gauge-links \\
\hline
incoming quark & 
$u_i(p)\,e^{ip\cdot\xi}$ &
$\langle X\vert\delta_{ij}\,\psi_j(\xi)\vert H\rangle\ e^{ip\cdot\xi}$ &
$(\,U^n_{[\xi;-\infty]}\,)_{ij}$\\
incoming antiquark & 
$\overline v_i(p)\,e^{ip\cdot\xi}$ &
$\langle X\vert\overline\psi_j(\xi)\,\delta_{ji}\vert H\rangle 
\ e^{ip\cdot\xi}$ &
$(\,U^n_{[-\infty;\xi]}\,)_{ji}$\\
incoming gluon &
$\epsilon_a(p)\,e^{ip\cdot\xi}$ &
$\langle X\vert\delta_{ab}\,F_b^{\mu\nu}(\xi)\vert H\rangle\ e^{ip\cdot\xi}$ &
$(\,U^n_{[\xi;-\infty]}\,)_{ab}$\\
outgoing quark & 
$\overline u_i(k)\,e^{-ik\cdot\xi}$ &
$\langle hX\vert\overline\psi_j(\xi)\,\delta_{ji}\vert 0\rangle
\ e^{-ik\cdot\xi}$ &
$(\,U^n_{[+\infty;\xi]}\,)_{ji}$\\
outgoing antiquark & 
$v_i(k)\,e^{-ik\cdot\xi}$ &
$\langle hX\vert\delta_{ij}\,\psi_j(\xi)\vert 0\rangle\ e^{-ik\cdot\xi}$ &
$(\,U^n_{[\xi;+\infty]}\,)_{ij}$\\
outgoing gluon &
$\epsilon^\ast_a(p)\,e^{-ik\cdot\xi}$ &
$\langle hX\vert F_b^{\mu\nu}(\xi)\,\delta_{ba}\vert 0\rangle\ e^{-ik\cdot\xi}$&
$(\,U^n_{[+\infty;\xi]}\,)_{ba}$\\[2pt]
\hline
\multicolumn{4}{c}{}\\
&&&contribution to\\[-2pt]
&`free' wave functions&fields in correlator&other gauge-links \\
\hline
incoming quark & 
$\overline u_i(p)$ &
$\langle H\vert\overline\psi_j(0)\,\delta_{ji}\vert X\rangle$ &
$(\,U^n_{[-\infty;0]}\,)_{ji}$\\
incoming antiquark & 
$v_i(p)$ &
$\langle H\vert\delta_{ij}\,\psi_j(0)\vert X\rangle $ &
$(\,U^n_{[0;-\infty]}\,)_{ij}$\\
incoming gluon &
$\epsilon_a(p)$ &
$\langle H\vert F_b^{\mu\nu}(0)\,\delta_{ba}\vert X\rangle$ &
$(\,U^n_{[-\infty;0]}\,)_{ba}$\\
outgoing quark & 
$u_i(k)$ &
$\langle 0\vert\delta_{ij}\,\psi_j(0)\vert hX\rangle$ &
$(\,U^n_{[0;+\infty]}\,)_{ij}$\\
outgoing antiquark & 
$\overline v_i(k)$ &
$\langle 0\vert\overline\psi_j(0)\,\delta_{ji}\vert hX\rangle$ &
$(\,U^n_{[+\infty;0]}\,)_{ji}$\\
outgoing gluon &
$\epsilon^\ast_a(p)$ &
$\langle 0\vert\delta_{ab}\,F_b^{\mu\nu}(0)\vert hX\rangle$ &
$(\,U^n_{[0;+\infty]}\,)_{ab}$\\[2pt]
\hline
\end{tabular}
\parbox{0.9\textwidth}{
\caption{Fields and gauge-links that play a role in hadron correlators. 
For comparison the standard free wave
functions of the partons in the hard scattering amplitudes are given in
the first column.
The second column shows how the partons appear in the correlators.
The third column gives the contributions of the external partons to the various gauge-links. 
The $U^n_{[a;b]}$ are the Wilson lines
along the light-cone direction $n$ given in Eq.~\eqref{WilsonLines}. 
The upper table is for the diagram corresponding to the hard
amplitude. The lower table for the diagram corresponding to the conjugate amplitude.
\label{TABLE1}}}
\end{table}

In order to find the gauge-link in a particular correlator one has to absorb the summation of all the gluon couplings to the hard part.
At leading twist this involves gluon fields collinear to the hadron's momentum $A^\mu\propto(A{\cdot}n)P^\mu$.
As will be argued in section~\ref{TAKAHASHI}, 
the resummation of all collinear gluon insertions leads to the attachment 
of Wilson lines to every external leg of the basic hard part, except to 
those that connect the hard part and the correlator under consideration.
The kind of Wilson lines attached to the external legs depends only on 
the nature of the external partons and not on the long or short distance 
processes. They are summarized in Table~\ref{TABLE1}.
The final result is, then, obtained by pulling all the Wilson lines 
through the color charges of the hard parts, 
moving them to the correlator under consideration. 
There they combine into the required gauge-link.
The whole subprocess-dependence of the gauge-link structure comes
from this last step. 

The resummation of collinear gluons described above has lead to a certain 
link structure in the $\xi^- \propto \xi{\cdot}P$ direction.
In the appendix we will indicate how this structure is appropriately closed in
the transverse direction at $\xi^-{=}{\pm}\infty$ by also 
taking transverse gluons into account.
This completes the derivation of the full gauge-link and
provides us with a calculational scheme (in an arbitrary gauge) for a general scattering diagram.
Summarizing, the calculational scheme consists of the following steps:
\begin{enumerate}
\item
Consider the diagram for a specific elementary squared amplitude.
Rather than as free spinors, the external partons appear as matrix elements 
given in the column `fields in correlator' in Table~\ref{TABLE1}, which
in turn appear in the distribution and fragmentation correlators. 
\item
In order to obtain the gauge-link in the TMD correlator of a specific
parton one needs to resum the insertions of gluons collinear to this parton.
For a particular correlator, this resummation is achieved by replacing 
the color wave functions of the other external partons with the appropriate 
Wilson lines. That is, the $\delta_{ij}$ for fermions and the 
$\delta_{ab}$ for vector bosons should be replaced by the Wilson lines given in the column `contribution to other gauge-links' in Table~\ref{TABLE1}. 
\item
Next one can use color flow identities, such as
\begin{equation}\label{ColorFlow}
t^a_{ij}t^a_{kl}
=T_F\Big(\,\delta_{il}\delta_{jk}
-\frac{1}{N_c}\,\delta_{ij}\delta_{kl}\,\Big)\ ,
\end{equation}
to pull the Wilson lines through the color structure of the hard parts.
Only the color structure of the hard part is relevant. 
All momentum dependence as well as all other correlators 
(in which the colors of partons in amplitude and conjugate amplitude are contracted)
can be discarded.
In the final expression the transverse pieces of the gauge-link must be included.
\item
The gauge-link is now simply obtained from the resulting expression.
By applying steps~1 through~3
one has obtained the expression for the bare diagram
with an additional structure multiplying the external leg of the parton 
under consideration.
This structure is the gauge-link that enters in the TMD correlation 
function of this parton.
\end{enumerate}
To illustrate the procedure outlined above, 
we will give some characteristic examples in the next section.

\section{examples\label{VOORBEELDEN}}

\subsection*{example A: distribution in SIDIS and DY}

\begin{figure}
\centering
\includegraphics{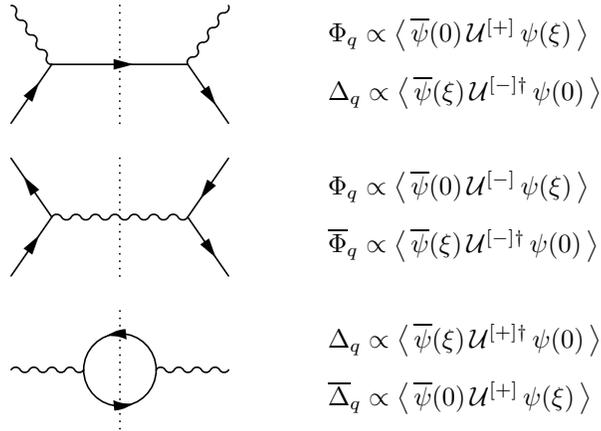}
\caption{The correlators for the leading hard
subprocesses in SIDIS, DY and $e^+e^-$-annihilation.\label{SidisDy}}
\end{figure}

Here we will show that the rules given in the previous section
reproduce the future and past pointing Wilson lines, equation~\eqref{GL},
appearing in the quark correlators in the SIDIS and DY processes.
The color parts of the expressions of the SIDIS, DY and $e^+e^-$-annihilation 
diagrams are $\tr[\Phi_q\Delta_q]$, $\tr[\Phi_q\overline\Phi_q]$ and
$\tr[\Delta_q\overline\Delta_q]$, respectively (see Fig.~\ref{SidisDy}).
Before the usual color summation (distribution) or color averaging
(fragmentation) is carried out, 
the matrix elements in the correlation functions are 
\begin{subequations}\label{MATRIXELEMENT1}
\begin{align}
\big(\Phi_q\big)_{ij}(p)
&=\sum_X\int\frac{\mathrm d^4\xi}{(2\pi)^4}\ e^{ip\cdot\xi}\,
\langle X\vert\,\psi_i(\xi)\,\vert H\rangle\,
\langle H\vert\,\overline\psi_j(0)\,\vert X\rangle\ ,\label{MATRIXELEMENT1c}\\
\big(\overline\Phi_q\big)_{ij}(p)
&=\sum_X\int\frac{\mathrm d^4\xi}{(2\pi)^4}\ e^{ip\cdot\xi}\,
\langle H\vert\,\psi_i(0)\,\vert X\rangle\,
\langle X\vert\,\overline\psi_j(\xi)\,\vert H\rangle\ ,\displaybreak[0]\\
\big(\Delta_q\big)_{ij}(p)
&=\sum_X\int\frac{\mathrm d^4\xi}{(2\pi)^4}\ e^{-ip\cdot\xi}\,
\langle 0\vert\,\psi_i(0)\,\vert hX\rangle\,
\langle hX\vert\,\overline\psi_j(\xi)\,\vert 0\rangle\ ,\label{ABC}\\
\big(\overline\Delta_q\big)_{ij}(p)
&=\sum_X\int\frac{\mathrm d^4\xi}{(2\pi)^4}\ e^{-ip\cdot\xi}\,
\langle hX\vert\,\psi_i(\xi)\,\vert 0\rangle\,
\langle 0\vert\,\overline\psi_j(0)\,\vert hX\rangle\ .
\end{align}
\end{subequations}
Following step~2, 
we resum all collinear interactions of gluons for the incoming quark
correlator $\Phi_q(p)$ in SIDIS and DY by taking the appropriate Wilson lines 
(as prescribed by Table~\ref{TABLE1})
instead of the matrix elements in Eq.~\eqref{MATRIXELEMENT1}
for the other correlators.
This amounts to making the replacement
\begin{subequations}
\begin{equation}\label{repl}
(\Delta_q)_{ij}
\rightarrow\big(U_{[0;\infty]}^nU_{[\infty;\xi]}^n\big)_{ij}\ ,
\end{equation}
in SIDIS and the replacement 
\begin{equation}
(\overline\Phi_q)_{ij}
\rightarrow\big(U_{[0;-\infty]}^nU_{[-\infty;\xi]}^n\big)_{ij}\ ,
\end{equation}
\end{subequations}
in DY.
The quark correlator $\Phi_q$ is left untouched 
since it is its gauge-link that we are calculating. 
Making these replacements and including the transverse pieces of the gauge-links at $\xi^-\propto\xi{\cdot}P={\pm}\infty$,
the color parts of the expressions for SIDIS and DY become
$\tr[\Phi_q\,\mathcal U^{[+]}]$ and $\tr[\Phi_q\,\mathcal U^{[-]}]$,
respectively.
For SIDIS we have now obtained the TMD correlator
(performing the integration over $p^-\propto p{\cdot}P$)
\begin{equation}
\Phi_q^{[+]}(x,p_{\sT})
=\int\frac{\mathrm d(\xi{\cdot}P)\mathrm d^2\xi_\sT}{(2\pi)^3}\
e^{ip\cdot\xi}\,\langle H|\,\overline\psi(0)\,\mathcal U^{[+]}
\,\psi(\xi)\,|H\rangle\ .
\end{equation}
Hence, we indeed reproduce the familiar gauge-invariant quark correlator 
with a future pointing Wilson line for SIDIS and a 
correlator with a past pointing Wilson line for DY, 
as indicated in Fig.~\ref{SidisDy}.
That figure enumerates all correlators appearing in simple electroweak processes.

The occurrence of a future pointing Wilson line in SIDIS and a past 
pointing Wilson line in DY is sometimes assigned to the virtual photon 
being space-like in SIDIS and time-like in DY.
Here we see that the appropriate point of view is that in 
SIDIS one gets a future pointing Wilson line, 
because the color flow runs via an outgoing quark.
Similarly, one gets a past pointing Wilson line in DY,
because there the color flow runs via an incoming antiquark.
That is the point of view that is generalized to arbitrary processes 
in this paper.

\subsection*{example B: fragmentation in SIDIS}

We consider the gauge-link that enters in the quark fragmentation correlator 
in SIDIS. This gauge-link is obtained by resumming all collinear gluons 
coming from the quark fragmentation correlator $\Delta(k)$.
Following step~2, 
this is done by attaching the appropriate Wilson lines to the matrix elements of the quark correlator~\eqref{MATRIXELEMENT1c} as prescribed by Table~\ref{TABLE1}.
This amounts to making the replacement
\begin{equation}\label{Replacement}
(\Phi_q)_{ij}
\rightarrow\big(U_{[\xi;-\infty]}^nU_{[-\infty;0]}^n\big)_{ij}\ .
\end{equation}
The fragmentation correlator $\Delta_q$ is left untouched,
since it is its gauge-link that we are calculating.
Making the replacement~\eqref{Replacement} in the color part 
$\tr[\Phi_q\Delta_q]$ of the expression for the SIDIS process,
it becomes $\tr[U_{[-\infty;0]}^n\Delta_q\,U_{[\xi;-\infty]}^n]$.
Hence, we have obtained the quark fragmentation correlator
(averaging over color indices is implicit)
\begin{equation}
\Delta_q^{[-]}(z,k_\st)
=\sum_X\frac{1}{4z}\int\frac{\mathrm{d}(\xi{\cdot}P_h)\,
\mathrm{d}^2 \xi_\st}{(2\pi)^3}\ e^{-ik{\cdot}\xi}\,
\langle 0\vert\,U_{[-\infty;0]}^n\psi(0)\,\vert hX\rangle\,
\langle hX\vert\,\overline\psi(\xi)U_{[\xi;-\infty]}^n\,\vert 0\rangle\ .
\end{equation}
The reason for writing the fragmentation correlator this way is the following.
When we say in Table~\ref{TABLE1} that the resummation of all collinear interactions amounts to attaching Wilson lines to the external legs of the diagram,
this applies to the \emph{color} structure of the Wilson lines.
However, the field operators $A$ in the gauge-link connecting to $\xi$ belong
to the matrix element $\langle hX|\overline\psi(\xi)|0\rangle$, while
the fields in the gauge-link connecting to the point $0$ belongs
to the matrix element $\langle 0|\psi(0)|hX\rangle$. 
A more detailed discussion on this point was given in Ref.~\cite{Bacchetta:2005rm}.
Certainly when considering proton-proton scattering,
where more complicated gauge-link structures will appear,
this becomes a very cumbersome notation.
In Fig.~\ref{SidisDy} we have symbolically written it as 
\begin{equation}\label{SYMBOLIC}
\Delta_q^{[-]}(z,k_\st)
=\sum_X\frac{1}{4z}\int\frac{\mathrm{d}(\xi{\cdot}P_h)\,
\mathrm{d}^2 \xi_\st}{(2\pi)^3}\ e^{-ik{\cdot}\xi}\,
\langle\,\overline\psi(\xi)\,\mathcal U^{[-]\dagger}\,\psi(0)\,\rangle\ ,
\end{equation}
and assume that it is understood that fields that connect to the point $\xi$ 
or $0$ appear in the appropriate matrix elements.
In the appendix we use similar symbolic notations for the quark and gluon fragmentation correlators.
In expression~\eqref{SYMBOLIC} we have also included the transverse 
pieces of the gauge-links.

\subsection*{example C: distribution in 
quark-quark scattering in proton-proton collisions}

\begin{figure}[t]
\centering
\begin{minipage}{3cm}
\centering
\includegraphics[width=3cm]{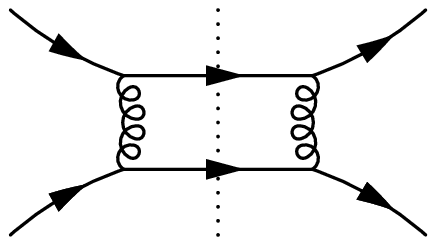}
\put(-85,13){$p_1$}
\put(-85,30){$p_2$}
\put(-55,5){$k_1$}
\put(-55,38){$k_2$}\\[2mm]
(a)
\end{minipage}\hspace{0.5cm}
\begin{minipage}{3cm}
\centering
\includegraphics[width=\textwidth]{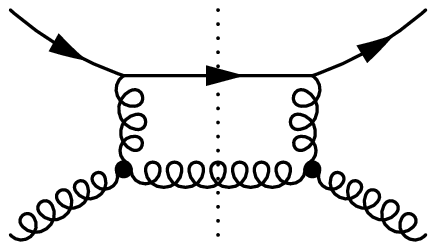}
\put(-85,13){$p_1$}
\put(-85,30){$p_2$}
\put(-55,0){$k_1$}
\put(-55,38){$k_2$}\\[2mm]
(b)
\end{minipage}\hspace{0.5cm}
\begin{minipage}{3cm}
\centering
\includegraphics[width=\textwidth]{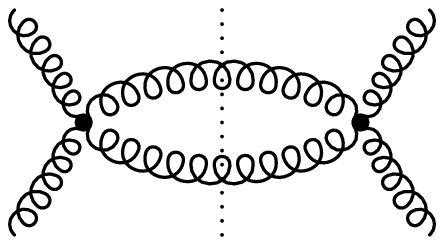}
\put(-90,13){$p_1$}
\put(-90,30){$p_2$}
\put(-55,3){$k_1$}
\put(-55,40){$k_2$}\\[2mm]
(c)
\end{minipage}
\parbox{0.73\textwidth}{
\caption{Three examples of partonic scattering processes that 
contribute in proton-proton collisions: 
(a) a quark-quark scattering contribution;
(b) a quark-gluon scattering contribution;
(c) a gluon-gluon scattering contribution.\label{contrib}}}
\end{figure}

The quark-quark scattering diagram in Fig.~\ref{contrib}a is one of the
partonic processes that contributes in proton-proton collisions.
In this example we will calculate the gauge-link that enters in the correlation function $\Phi_q(p_1)$ of the quark in the lower-left corner.
This gauge-link was already calculated in Ref.~\cite{Bacchetta:2005rm}
(c.f.\ Eq.~(A10-A11)).
The expression for the diagram in Fig.~\ref{contrib}a is proportional to
the color-traced expression
\begin{equation}\label{EXPRfig3a}
\tr\big[\,\Phi_q(p_1)t^a\Delta_q(k_1)t^b\,\big]
\tr\big[\,\Phi_q(p_2)t^a\Delta_q(k_2)t^b\,\big]\ .
\end{equation}
Following step~2, we resum all interactions of collinear gluons coming from $\Phi_q(p_1)$ by attaching the appropriate Wilson lines to all the matrix elements in $\Phi_q(p_2)$ and $\Delta_q(k_i)$ 
(see~\eqref{MATRIXELEMENT1c} and~\eqref{ABC})
as prescribed by Table~\ref{TABLE1}.
This amounts to making the replacements~\eqref{Replacement} and~\eqref{repl} for the quark correlator $\Phi_q(p_2)$ and fragmentation correlators
$\Delta_q(k_i)$, respectively.
For the same reason as in example~A the correlator $\Phi_q(p_1)$ is left untouched.
With these substitutions the expression~\eqref{EXPRfig3a} becomes,
including the transverse pieces of the gauge-links,
\begin{equation}\label{ExampleIa}
\tr\big[\,\Phi_q(p_1)t^a\mathcal U^{[+]}t^b\,\big]
\tr\big[\,\mathcal U^{[-]\dagger}t^a\mathcal U^{[+]}t^b\,\big]\ .
\end{equation}
As it stands in Eq.~\eqref{ExampleIa}, 
the future and past pointing Wilson lines cannot be absorbed in the correlator $\Phi_q(p_1)$.
However, using the color flow identity~\eqref{ColorFlow} it can be rewritten to
\begin{equation}
\tr\Big[\,\Phi_q(p_1)\,
\Big(\,\frac{N_c^2{+}1}{N_c^2{-}1}\;\mathcal U^{[+]}
\frac{\tr\big[\,\mathcal U^{[+]}\mathcal U^{[-]\dagger}\,\big]}{N_c}
-\frac{2}{N_c^2{-}1}\;\mathcal U^{[+]}\mathcal U^{[-]\dagger}
\mathcal U^{[+]}\,\Big)\,t^at^b\,\Big]\,
\tr\big[\,t^at^b\,\big]\ .
\end{equation}
In this expression all link structures multiply the correlator $\Phi_q(p_1)$, 
as required by step~3.
The structure multiplying the correlator can now be absorbed in it.
What remains is the expression for the tree-level diagram
$\tr\big[\Phi_q^{[\mathcal U]}(p_1)t^at^b\big]
\tr\big[t^at^b\big]$
involving the gauge-invariant correlator 
\begin{equation}
\Phi_q^{[\mathcal U]}(x_1,p_{1\sT})
=\int\frac{\mathrm d(\xi{\cdot}P)\mathrm d^2\xi_\sT}{(2\pi)^3}\ 
e^{ip_1\cdot\xi}\,\langle H|\,\overline\psi(0)\,
\Big\{\,\frac{N_c^2{+}1}{N_c^2{-}1}\,
\frac{\tr\big(\mathcal U^{[\Box]}\big)}{N_c}\mathcal U^{[+]}
-\frac{2}{N_c^2{-}1}\,\mathcal U^{[\Box]}\mathcal U^{[+]}\,\Big\}\,
\psi(\xi)\,|H\rangle\ .
\end{equation}
This agrees with the result obtained in Eq.~(A11a) in Ref.~\cite{Bacchetta:2005rm}.
The gauge-invariant correlators for the other quark-quark scattering channels are enumerated in Table~\ref{Tqq2qq} in the appendix.

\subsection*{example D: distribution in
quark-gluon scattering in proton-proton collisions}

As our fourth example we will consider the diagram in Fig.~\ref{contrib}b,
which is one of the possible quark-gluon scattering diagrams contributing
in proton-proton collisions. 
We will calculate the gauge-link structure
in the correlation function for the incoming gluon in the lower-left corner.
The gluon correlators contain gluon fields involving,
at leading order, transverse indices and they can be expressed in terms
of the field strength tensor. 
The matrix elements in the correlators
(omitting the Lorentz indices on the fields) are~\cite{Ji:1992eu,Jaffe:1995an,Bashinsky:1998if,Mulders:2000sh,Ji:2005nu}
\begin{subequations}\label{MATRIXELEMENT2}
\begin{equation}\label{MATRIXELEMENT2a}
\big(\Phi_g\big)_{ab}(p)
=\sum_X\int\frac{\mathrm d^4\xi}{(2\pi)^4}\ e^{ip\cdot\xi}\,
\langle X|\,F_a(\xi)\,|H\rangle\,
\langle H|\,F_b(0)\,|X\rangle\ ,
\end{equation}
for the gluon distribution correlator and
\begin{equation}
\big(\Delta_g\big)_{ab}(k)
=\sum_X\int\frac{\mathrm d^4\xi}{(2\pi)^4}\ e^{-ik\cdot\xi}\,
\langle 0|\,F_a(0)\,|hX\rangle\,
\langle hX|\,F_b(\xi)\,|0\rangle\ ,
\end{equation}
\end{subequations}
for the gluon fragmentation correlator. 
The color part of the expression for the diagram in Fig.~\ref{contrib}b is 
\begin{equation}\label{EXPRfig3b}
\tr\big[\,\Phi_q(p_2)t^a\Delta_q(k_2)t^b\,\big]\,if^{acd}\,if^{bc'd'}\,
\big(\Delta_g\big)_{cc'}(k_1)\,\big(\Phi_g\big)_{d'd}(p_1)\ .
\end{equation}
Following step~2, 
we resum all interactions of collinear gluons for $\Phi_g(p_1)$ by attaching the appropriate Wilson lines to all the matrix elements in~\eqref{MATRIXELEMENT1} and~\eqref{MATRIXELEMENT2} as prescribed by Table~\ref{TABLE1}.
This amounts to making the replacements~\eqref{Replacement} and~\eqref{repl} for the quark distribution and fragmentation correlators, respectively,
and the replacement
\begin{equation}
\big(\Delta_g\big)_{ab}(k_1)
\rightarrow\big(U_{[0;\infty]}^nU_{[\infty;\xi]}^n\big)_{ab}\ ,
\end{equation}
for the gluon fragmentation correlator.
For the same reason as in example~A the gluon correlator $\Phi_g(p_1)$ is left untouched.
Expression~\eqref{EXPRfig3b} now becomes,
including the transverse pieces of the gauge-links,
\begin{align*}
\tr\big[\,&\mathcal U^{[-]\dagger}t^a\mathcal U^{[+]}t^b\,\big]\,
if^{acd}\,if^{bc'd'}\,
\mathcal U_{cc'}^{[+]}\,\big(\Phi_g\big)_{d'd}(p_1)\\
&=\frac{1}{T_F}\tr\big[\,\mathcal U^{[-]\dagger}t^a\mathcal U^{[+]}t^b\,\big]\,
\tr\big[\,[t^a,t^d]\mathcal U^{[+]}[t^b,t^{d'}]\mathcal U^{[+]\dagger}\,\big]\,
\big(\Phi_g\big)_{d'd}(p_1)\\
&=-2T_FN_c\tr\Big[\,t^d\mathcal U^{[+]}t^{d'}
\Big(\tfrac{1}{2}\mathcal U^{[-]\dagger}
+\tfrac{1}{2}\frac{\tr[\mathcal U^{[\Box]}]}{N_c}\mathcal U^{[+]\dagger}\Big)\,
\Big]\,\big(\Phi_g\big)_{d'd}(p_1)\ ,
\end{align*}
where we have used~\eqref{ADJrep} in the first step and the color flow 
identity~\eqref{ColorFlow} in the second step. 
Inserting the expression for the gluon correlator~\eqref{MATRIXELEMENT2a},
we see that the expression for the gauge-invariant gluon correlator corresponding 
to the incoming gluon in Fig.~\ref{contrib}b is
\begin{equation}\begin{split}\label{XYZ}
\Phi_g^{[\mathcal U]}(x_1,p_{1\sT})
=\int\frac{\mathrm d(\xi{\cdot}P)\mathrm d^2\xi_\sT}{(2\pi)^3}\ 
e^{ip_1\cdot\xi}\,\langle H|\tr\big[\,F(0)\,\mathcal U^{[+]}\,F(\xi)\,
\Big\{\tfrac{1}{2}\mathcal U^{[-]\dagger}
+\tfrac{1}{2}\frac{\tr[\mathcal U^{[\Box]}]}{N_c}
\mathcal U^{[+]\dagger}\Big\}\,\big]\,|H\rangle\ .
\end{split}\end{equation}
The factor $-2T_FN_c$ was not included in the gluon correlator,
since this is the color factor of the tree-level diagram.
The gauge-links for the other quark-gluon scattering channels are enumerated in Table~\ref{Tqg2qg} in the appendix.

\subsection*{example E: distribution in
gluon-gluon scattering in proton-proton collisions}

As said, the gauge-link is obtained by pulling the color structure of the Wilson lines attached to all the external legs through the hard part.
Therefore, it was only necessary to take the color structure of the hard parts into account in the previous examples.
That is allowed in those examples,
because the color structure of the hard parts can be factored from the rest of the diagram.
For gluon-gluon scattering processes involving only three-point vertices 
this is also the case.
When considering diagrams containing 4-gluon vertices the color structure 
is more complex. The vertex is given by 
\begin{equation*}
\qquad
\parbox{0.1\textwidth}{
\includegraphics[width=0.1\textwidth]{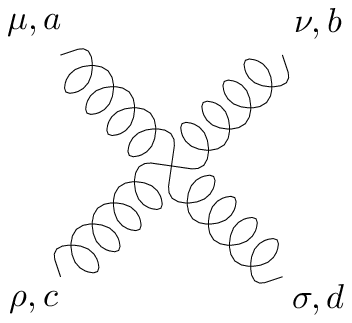}}
\quad
=-ig^2\,\big(\,f^{abe}f^{cde}\,g^{\mu[\rho}g^{\sigma]\nu}
+f^{ace}f^{bde}\,g^{\mu[\nu}g^{\sigma]\rho}
+f^{ade}f^{bce}\,g^{\mu[\nu}g^{\rho]\sigma}\,\big)\ .
\end{equation*}
However, a diagram containing such a 4-gluon vertex can be written as a sum of terms,
with a multiplicative color structure for each term separately.
The color structure of these terms correspond to the color structure of 
one of the diagrams with only 3-point vertices in Table~\ref{Tgg2gg}.
The terms will, then, 
get the corresponding gauge-link multiplied by the color factor of that diagram.
As an explicit example we take the diagram in Fig.~\ref{contrib}c
\begin{equation}\label{VIER}
\parbox{2.5cm}{
\includegraphics[width=2.5cm]{Figures/gg2ggC.eps}
\put(-82,0){$a_1$}
\put(2,0){$a_2$}
\put(-82,36){$a_3$}
\put(2,36){$a_4$}
\put(-33,4){$a_5$}
\put(-47,4){$a_6$}
\put(-33,32){$a_7$}
\put(-47,32){$a_8$}}
\qquad\quad\begin{split}
\propto\ &\big(\,c_1\,f^{a_1a_6b}f^{a_3a_8b}
+c_2\,f^{a_1a_3b}f^{a_6a_8b}
+c_3\,f^{a_1a_8b}f^{a_3a_6b}\,\big)\\
&\times\big(\,c_1\,f^{a_2a_5c}f^{a_4a_7c}
+c_2\,f^{a_2a_4c}f^{a_5a_7c}
+c_3\,f^{a_2a_7c}f^{a_4a_5c}\,\big)\ ,
\end{split}
\end{equation}
where the $c_i$ are functions of the kinematical variables of the process.
The r.h.s.\ can be represented by
\begin{equation}\begin{split}\label{FETZEPIJLMAN}
c_1&c_1\ \parbox{1.5cm}{\includegraphics[width=1.5cm]{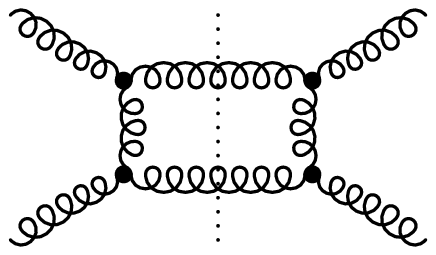}}\,
+c_2c_2\ \parbox{1.5cm}{\includegraphics[width=1.5cm]{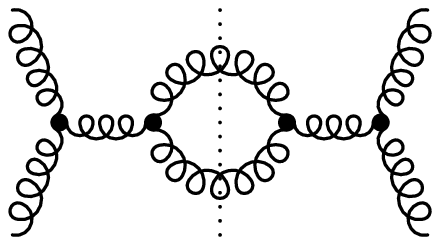}}\,
+c_3c_3\ \parbox{1.5cm}{\includegraphics[width=1.5cm]{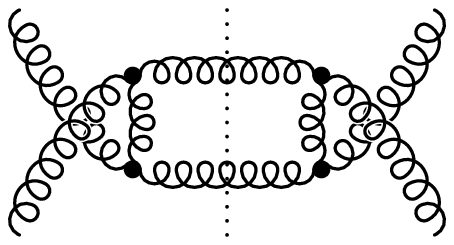}}
+c_1c_2\; \bigg(\;\parbox{1.5cm}{\includegraphics[width=1.5cm]{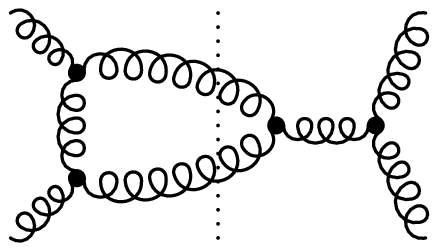}}\,
+\,\parbox{1.5cm}{\includegraphics[width=1.5cm]{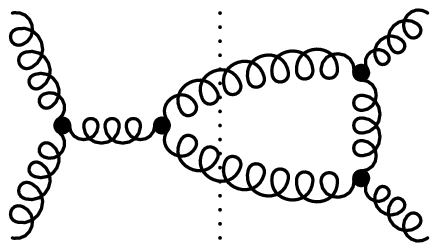}}\;\bigg)\,\\
&+c_2c_3\;
\bigg(\;\parbox{1.5cm}{\includegraphics[width=1.5cm]{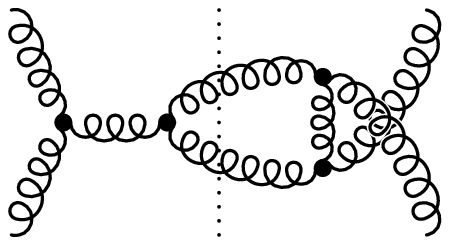}}\,
+\,\parbox{1.5cm}{\includegraphics[width=1.5cm]{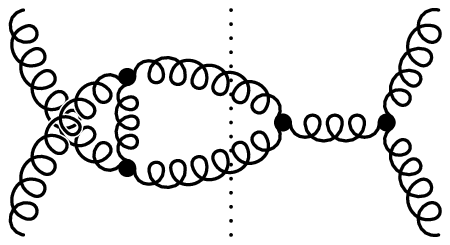}}\;\bigg)\,
+c_1c_3\; \bigg(\;\parbox{1.5cm}{\includegraphics[width=1.5cm]{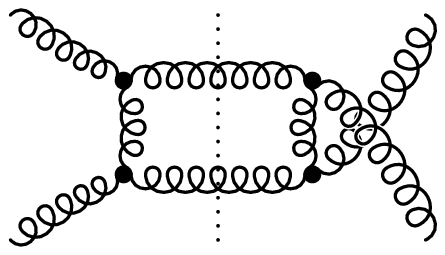}}\,
+\,\parbox{1.5cm}{\includegraphics[width=1.5cm]{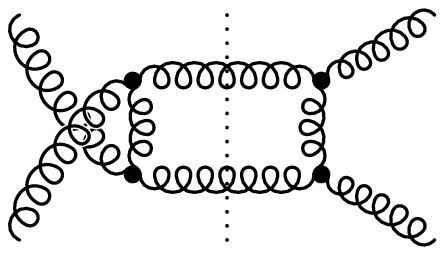}}\;\bigg)\ .
\end{split}\end{equation}
Diagrams in this expression only depict the \emph{color part} of that diagram.
The expression for Fig.~\ref{contrib}c is obtained from~\eqref{FETZEPIJLMAN} 
by replacing the diagrams by their color factors times their gauge-links,
as they can be read from Table~\ref{Tgg2gg}.
Note that not each term appearing in the expression above appears in that Table,
such as, for instance, the third structure in~\eqref{FETZEPIJLMAN}.
This term is related to the first by interchanging the two outgoing particles.
Therefore, its gauge-link structure is identical to that term.
Similarly, diagrams that are simply related to one of the diagrams in 
Table~\ref{Tgg2gg} by interchanging the two outgoing particles were not 
included in that Table, since the gauge-link structures are identical.
With these remarks in mind
we find that the color part of the expression for Fig.~\ref{contrib}c is
\begin{equation}\begin{split}
\Phi_g^{[1]}&(p_1)\Phi_g^{[1]}(p_2)\big[\,4T_FN_c^2(c_1c_1{+}c_3c_3)\,\big]
\Delta_g^{[1]}(k_1)\Delta_g^{[1]}(k_2)\\
&+\Phi_g^{[2]}(p_1)\Phi_g^{[2]}(p_2)\big[\,4T_FN_c^2(c_2c_2{+}c_1c_2{-}c_2c_3)\,\big]
\Delta_g^{[2]}(k_1)\Delta_g^{[2]}(k_2)\\
&+\Phi_g^{[3]}(p_1)\Phi_g^{[3]}(p_2)\big[\,4T_FN_c^2c_1c_3\,\big]
\Delta_g^{[3]}(k_1)\Delta_g^{[3]}(k_2)\ ,
\end{split}\end{equation}
where $\Phi_g^{[1]}$, $\Phi_g^{[2]}$ and $\Phi_g^{[3]}$ are the first, second 
and third distribution correlators in Table~\ref{Tgg2gg}, respectively 
(and similarly for the fragmentation correlators $\Delta_g^{[i]}$).

\section{Resummation of collinear gluons into gauge-links\label{TAKAHASHI}}

\begin{figure}
\includegraphics{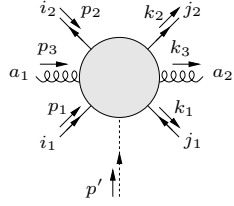}
\caption{The truncated Green's function
$\Gamma_{(i_2j_2)(i_1j_1)}^{a_2a_1}(p',p_1,p_2,p_3,k_1,k_2,k_3)$.
\label{qqg2qqg}}
\end{figure}

In this section we will give some leading twist arguments to argue the correctness of the conjectures made in the previous sections.
These arguments can possibly be generalized to subleading twist using the methodology of~\cite{Boer:1999si}.
We will take the partonic process 
$\phi(p')q(p_1)\bar q(p_2)g(p_3){\rightarrow}\bar q(k_1)q(k_2)g(k_3)$ 
as an example, Fig.~\ref{qqg2qqg}.
We do not specify what kind of parton $\phi(p')$ is,
since that is not important at this stage (hence the dashed line).
This process is only schematic, 
chosen such that it contains all types of partons in the initial and in the final state.
Therefore,
our results are straightforwardly generalized to arbitrary partonic processes.

\begin{figure}
\begin{minipage}[t]{3cm}
\centering
\includegraphics[width=\textwidth]{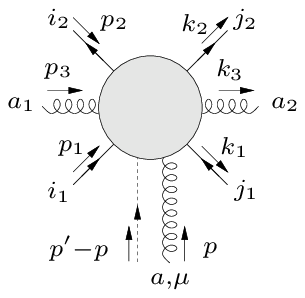}\\
(a)
\end{minipage}
\hspace{.5cm}
\begin{minipage}[t]{3cm}
\centering
\includegraphics[width=\textwidth]{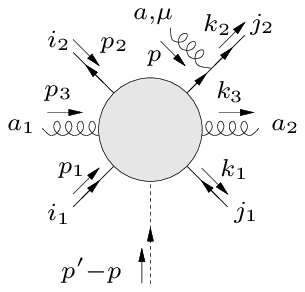}\\
(b)
\end{minipage}
\parbox{0.9\textwidth}{\caption{
Two possible one-gluon insertions where the gluon momentum $p$ is collinear to $p^\prime$:
(a) the insertion of the additional collinear gluon in the truncated amplitude involving the truncated Green's function
$\Gamma_{(i_2j_2)(i_1j_1)}^{(\mu a)a_2a_1}
(p;p'{-}p,p_1,p_3,p_3,k_1,k_2,k_3)$;
(b) the insertion of the additional collinear gluon to the outgoing quark.
\label{qqg2qqgG}}}
\end{figure}

In a hadronic scattering process the diagram in Fig.~\ref{qqg2qqg} is described by a truncated Green's function 
$\Gamma_{(i_2j_2)(i_1j_1)}^{a_2a_1}$ 
with (approximately) on mass-shell external momenta.
The external legs of this Green's function are connected to correlators
(or possibly to cut propagators), that is
$\big(\Phi_q\big)_{i_1k}(p_1)$, $\big(\overline\Phi_q\big)_{li_2}(p_2)$, 
$\big(\Phi_g\big)_{a_1b}(p_3)$, $\big(\overline\Delta_q\big)_{j_1m}(k_1)$,
$\big(\Delta_q\big)_{nj_2}(k_2)$ and $\big(\Delta_g\big)_{ca_2}(k_3)$.
Indices $i$ and $j$ will denote color-triplet indices
and indices $a$ and $b$ will be color-octet indices.
Here we will calculate the gauge-link that enters in the correlator $\Phi(p')$ corresponding to the parton $\phi(p')$.
It is determined by resumming all insertions of collinear gluons coming from the same hadron as $\phi(p')$.
Two examples of single-gluon insertions are depicted in Fig.~\ref{qqg2qqgG}.
The gluon $p$ couples everywhere except to the external leg corresponding to $\phi(p'{-}p)$. 
This parton involves the generic correlator 
\begin{equation}\label{QQAcorr}
\Phi^{a\mu}(p'{-}p,p)
\propto
\int\frac{\mathrm d^4\xi}{(2\pi)^4}\frac{\mathrm d^4\eta}{(2\pi)^4}\ 
e^{i(p'-p)\cdot\xi}e^{ip\cdot\eta}\,
\langle P|\,\phi^\dagger(0)\,A^{a\mu}(\eta)\,\phi(\xi)\,|P\rangle\ .
\end{equation}
To leading twist it is collinear to the parent hadron
$\Phi^\mu(p'{-}p,p)\propto P^\mu$.
The same holds true for the corresponding partons $p{=}xP$ and
$p'{=}x'P$.
Therefore, the expressions in Fig.~\ref{qqg2qqgG} are effectively contracted with $P^\mu$ 
(compare this with the Ward-Takahashi identity where the hard process in contracted with $p^\mu{=}xP^\mu$).

\begin{figure}
\begin{minipage}[t]{5cm}
\centering
\begin{alignat*}{2}
\parbox{2.5cm}{\includegraphics[width=2.5cm]{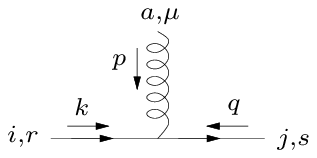}}\ 
&=ig\,V^\mu_{sr}(p,k,q)\,(T^a)_{ji}&\qquad
&\left\{\begin{array}{l}
V^\mu_{sr}=\gamma^\mu_{sr}\\[2mm]
(T^a)_{ji}=t^a_{ji}
\end{array}\right.\\[3mm]
\parbox{2.5cm}{\includegraphics[width=2.5cm]{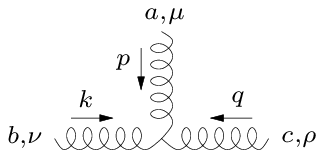}}\ 
&=ig\,W^\mu_{\nu\rho}(p,k,q)\,(T^a)_{bc}&
&\left\{\begin{array}{l}
W^\mu_{\nu\rho}
=g^{\mu}_{\phantom{\mu}\nu}(p{-}k)_\rho
{+}g_{\nu\rho}(k{-}q)^\mu{+}g^{\mu}_{\phantom{\mu}\rho}(q{-}p)_\nu\\[2mm]
(T^a)_{bc}={-}if^{abc}
\end{array}\right.
\end{alignat*}
\end{minipage}
\caption{3-point vertices in QCD.\label{3PointVertices}}
\end{figure}

We will start by considering all interactions of the collinear gluon $p$ with external partons
(except the one with $\phi(p')$).
These interactions all involve 3-point vertices that, 
in QCD, 
can be factored in a vector and a color part, 
see Fig.~\ref{3PointVertices}.
We use $T^a$ to denote any representation of the color matrices,
such that this is the fundamental or defining representation whenever we use color-triplet indices $(T^a)_{ij}{=}t^a_{ij}$ and the adjoint representation whenever we use color-octet indices $(T^a)_{bc}{=}{-}if^{abc}$.
For instance, the quark-quark-gluon vertex is
$igV^\mu(T^a)_{ji}{=}ig\gamma^\mu t^a_{ji}$.
With this notation the sum of all interactions with the external partons is
(here and in the rest of this section we suppress all Dirac and Lorentz indices, except the one that is contracted with the hadron momentum $P$)
\begin{subequations}\label{legs}
\begin{alignat}{2}
P_\mu&\;igV^\mu(T^a)_{j_2j'}\;S_{j'j}(k_2{-}p)\;
\Gamma_{(i_2j)(i_1j_1)}^{a_2a_1}
(p'{-}p,p_1,p_2,p_3,k_1,k_2{-}p,k_3)
&&\text{\small(outgoing quark)}\label{legsA}\\ 
&
+P_\mu\;igV^\mu(T^a)_{i_2i'}\;S_{i'i}\big({-}(p_2{+}p)\big)\;
\Gamma_{(ij_2)(i_1j_1)}^{a_2a_1}
(p'{-}p,p_1,p_2{+}p,p_3,k_1,k_2,k_3)
&&\text{\small(incoming antiquark)}\label{legsB}\\ 
&
+P_\mu\;igW^\mu(T^a)_{a_2b'}\;D_{b'b}(k_3{-}p)\;
\Gamma_{(i_2j_2)(i_1j_1)}^{ba_1}
(p'{-}p,p_1,p_2,p_3,k_1,k_2,k_3{-}p)
&&\text{\small(outgoing gluon)}\label{legsC}\displaybreak[0]\\ 
&+
\Gamma_{(i_2j_2)(ij_1)}^{a_2a_1}
(p'{-}p,p_1{+}p,p_2,p_3,k_1,k_2,k_3)\;
S_{ii'}(p_1{+}p)\;P_\mu\;igV^\mu(T^a)_{i'i_1}
&&\text{\small(incoming quark)}\label{legsD}\\ 
&+
\Gamma_{(i_2j_2)(i_1j)}^{a_2a_1}
(p'{-}p,p_1,p_2,p_3,k_1{-}p,k_2,k_3)\;
S_{jj'}\big({-}(k_1{-}p)\big)\;P_\mu\;igV^\mu(T^a)_{j'j_1}
&&\text{\small(outgoing antiquark)}\label{legsE}\\ 
&+
\Gamma_{(i_2j_2)(i_1j_1)}^{a_2b}
(p'{-}p,p_1,p_2,p_3{+}p,k_1,k_2,k_3)\;
D_{bb'}(p_3{+}p)\;P_\mu\;igW^\mu(T^a)_{b'a_1}\ .
&\mspace{11mu}&\text{\small(incoming gluon)}\label{legsF}
\end{alignat}
\end{subequations}
The propagators $S$ and $D$ are quark and gluon propagators respectively
(for convenience we use a time-like axial gauge).
With the notations defined above we see that the term~\eqref{legsA} represents the diagram where the collinear gluon $p$ couples to the outgoing quark $k_2$,
Fig.~\ref{qqg2qqgG}b.
In this case the vertex and the additional propagator are
\begin{equation}\label{VertexProp}
P_\mu\;igV^\mu(T^a)_{j_2j'}\;S_{j'j}(k_2{-}p)
=ig\slash P t_{j_2j'}^a\;
\delta_{j'j}\frac{i(\slash k_2{-}\slash p)}{(k_2{-}p)^2{+}i\epsilon}\ .
\end{equation}
As said at the beginning of this section,
this expression will appear left-multiplied by the quark fragmentation correlator $\Delta(k_2)$.
The product $\Delta(k_2)\,\slash k_2$ will only contribute at $\mathcal O(1/Q)$
and can be neglected in our leading twist discussion.
Therefore~\eqref{VertexProp} effectively equals
\begin{equation}\label{MultiplicativeFactor}
P_\mu\;igV^\mu(T^a)_{j_2j'}\;S_{j'j}(k_2{-}p)
=\frac{g(T^a)_{j_2j}}{x-i\epsilon}\ ,
\end{equation}
where we have also used that at leading twist all the external momenta can be treated on mass-shell $p_i^2{=}k_i^2{=}P^2{=}0$.
With the effective form~\eqref{MultiplicativeFactor} the term corresponding to the outgoing quark $k_2$ becomes
\begin{equation}\begin{split}\label{WaarschijnlijkWel}
\frac{g(T^a)_{j_2j}}{x-i\epsilon}\,&
\Gamma_{(i_2j)(i_1j_1)}^{a_2a_1}(p',p_1,p_2,p_3,k_1,k_2,k_3)\\
&+\frac{g(T^a)_{j_2j}}{x-i\epsilon}\,\Big\{\,
\Gamma_{(i_2j)(i_1j_1)}^{a_2a_1}
(p'{-}p,p_1,p_2,p_3,k_1,k_2{-}p,k_3)
-\Gamma_{(i_2j)(i_1j_1)}^{a_2a_1}(p',p_1,p_2,p_3,k_1,k_2,k_3)\,
\Big\}\ ,
\end{split}\end{equation}
where we have added and subtracted a term corresponding to a zero-momentum gluon $p^\mu{=}0^\mu$.
The second line of~\eqref{WaarschijnlijkWel} is actually pole independent,
because the residue of the expression between braces $\{\,\cdots\,\}$ vanishes at the pole $x{=}0$.
Therefore, 
the $(x{-}i\epsilon)^{-1}$ in the second line may be replaced by any other pole prescription.
All other external legs can be handled similarly,
for gluons using that at leading twist $k_\nu\Phi_g^{\nu\rho}(k)$ and the gauge-dependent terms do not contribute.

\begin{figure}
\centering
\includegraphics[width=0.9\textwidth]{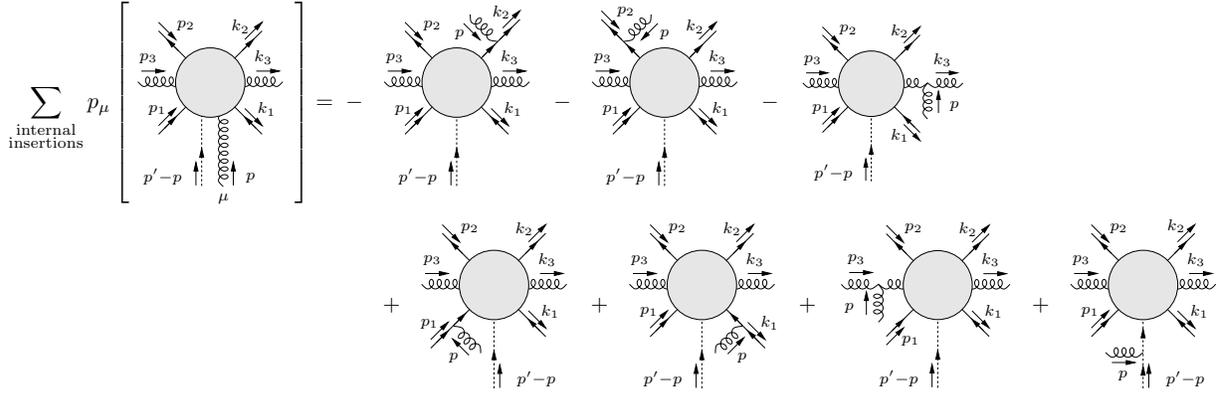}
\caption{Ward-Takahashi identity for truncated Green's functions in QCD.\label{SINTERKLAAS}}
\end{figure}

Next we will consider the internal interactions of the collinear gluon with the Green's function.
The sum of all internal insertions can be obtained from a Ward-Takahashi identity for truncated Green's functions in the axial gauge
(making the same suppression of indices as was mentioned earlier)
\begin{subequations}\label{WaTa}
\begin{alignat}{2}
\sum_{\substack{\text{internal}\\ \text{insertions}}}
&p_\mu\,
\Gamma_{(i_2j_2)(i_1j_1)}^{(\mu a)a_2a_1}
(p;p'{-}p,p_1,p_2,p_3,k_1,k_2,k_3)
\displaybreak[0]\nonumber\\
&=-g(T^a)_{j_2j}\,
\Gamma_{(i_2j)(i_1j_1)}^{a_2a_1}
(p'{-}p,p_1,p_2,p_3,k_1,k_2{-}p,k_3)
&&\text{(outgoing quark)}\label{contrib-2}\\ 
&\qquad-g(T^a)_{i_2i}\,
\Gamma_{(ij_2)(i_1j_1)}^{a_2a_1}
(p'{-}p,p_1,p_2{+}p,p_3,k_1,k_2,k_3)
&&\text{(incoming antiquark)}\\ 
&\qquad-g(T^a)_{a_2b}\,
\Gamma_{(i_2j_2)(i_1j_1)}^{ba_1}
(p'{-}p,p_1,p_2,p_3,k_1,k_2,k_3{-}p)
&&\text{(outgoing gluon)}\displaybreak[0]\\ 
&\qquad
+\Gamma_{(i_2j_2)(ij_1)}^{a_2a_1}
(p'{-}p,p_1{+}p,p_2,p_3,k_1,k_2,k_3)\,g(T^a)_{ii_1}
&&\text{(incoming quark)}\\ 
&\qquad
+\Gamma_{(i_2j_2)(i_1j)}^{a_2a_1}
(p'{-}p,p_1,p_2,p_3,k_1{-}p,k_2,k_3)\,g(T^a)_{jj_1}
&&\text{(outgoing antiquark)}\\ 
&\qquad
+\Gamma_{(i_2j_2)(i_1j_1)}^{a_2b}
(p'{-}p,p_1,p_2,p_3{+}p,k_1,k_2,k_3)\,g(T^a)_{ba_1}
&&\text{(incoming gluon)}\\ 
&\qquad
+\Gamma_{(i_2j_2)(i_1j_1)}^{a_2a_1}
(p',p_1,p_2,p_3,k_1,k_2,k_3)\,gT^a(\phi)\ .
&\mspace{100mu}&\text{(incoming $\phi(p')$)}\label{contrib-1}
\end{alignat}
\end{subequations}
(The representation $T^a(\phi)$ of the color matrices depends on the nature of parton $\phi(p')$).
This is pictorially represented in Fig.~\ref{SINTERKLAAS} where the blobs represent truncated Green's functions and where one should only include the color factors of the other perturbative elements.
In our application we contract the expression with the vector $P^\mu$ rather than with 
$p^\mu{=}xP^\mu$.
That expression can be obtained from~\eqref{WaTa} by dividing out the momentum fraction $x$.
In doing so one is creating a new pole at $x{=}0$ that needs to be treated with a certain pole prescription.
We will write $[x^{-1}]$ to indicate any pole prescription.
It is irrelevant which prescription is taken due to the following observation.
Dividing~\eqref{WaTa} by $x$ we see that~\eqref{contrib-2} cancels the first term in the second line of~\eqref{WaarschijnlijkWel} whichever prescription was taken,
since that term was independent of the pole prescription.
There is a similar cancellation between all terms corresponding to the same external partons in~\eqref{legs} and~\eqref{WaTa},
except for parton $\phi(p')$.
The total result for 
all internal and leg-insertions (except to $\phi(p')$) of the collinear gluon $p$ with polarization along $P$, 
then, becomes
\begin{subequations}\label{insertions-result}
\begin{alignat}{2}
\sum_{\text{insertions}}
&P_\mu\,\Gamma_{(i_2j_2)(i_1j_1)}^{(\mu a)a_2a_1}
(p;p'{-}p,p_1,p_2,p_3,k_1,k_2,k_3)\displaybreak[0]\nonumber\\
&=\frac{g(T^a)_{j_2j}}{x-i\epsilon}\,
\Gamma_{(i_2j)(i_1j_1)}^{a_2a_1}(p',p_i,k_i)
&\mspace{150mu}&\text{(outgoing quark)}\\
&\qquad
+\frac{g(T^a)_{i_2i}}{x+i\epsilon}\,
\Gamma_{(ij_2)(i_1j_1)}^{a_2a_1}(p',p_i,k_i)
&\mspace{150mu}&\text{(incoming antiquark)}\\
&\qquad
+\frac{g(T^a)_{a_2b}}{x-i\epsilon}\,
\Gamma_{(i_2j_2)(i_1j_1)}^{ba_1}(p',p_i,k_i)
&\mspace{150mu}&\text{(outgoing gluon)}\displaybreak[0]\\
&\qquad
-\Gamma_{(i_2j_2)(ij_1)}^{a_2a_1}(p',p_i,k_i)\,
\frac{g(T^a)_{ii_1}}{x+i\epsilon}
&\mspace{150mu}&\text{(incoming quark)}\\
&\qquad
-\Gamma_{(i_2j_2)(i_1j)}^{a_2a_1}(p',p_i,k_i)\,
\frac{g(T^a)_{jj_1}}{x-i\epsilon}
&\mspace{150mu}&\text{(outgoing antiquark)}\\
&\qquad
-\Gamma_{(i_2j_2)(i_1j_1)}^{a_2b}(p',p_i,k_i)\,
\frac{g(T^a)_{ba_1}}{x+i\epsilon}
&\mspace{150mu}&\text{(incoming gluon)}\displaybreak[0]\\
\lefteqn{\mspace{60mu}
-g\Big[\frac{1}{x}\Big]\,\Big\{\,
(T^a)_{j_2j}\,\Gamma_{(i_2j)(i_1j_1)}^{a_2a_1}(p',p_i,k_i)
+(T^a)_{i_2i}\,\Gamma_{(ij_2)(i_1j_1)}^{a_2a_1}(p',p_i,k_i)}\nonumber\\
\lefteqn{\mspace{150mu}
+(T^a)_{a_2b}\,\Gamma_{(i_2j_2)(i_1j_1)}^{ba_1}(p',p_i,k_i)
-\Gamma_{(i_2j_2)(ij_1)}^{a_2a_1}(p',p_i,k_i)(T^a)_{ii_1}}\nonumber\\
\lefteqn{\mspace{150mu}
-\Gamma_{(i_2j_2)(i_1j)}^{a_2a_1}(p',p_i,k_i)\,(T^a)_{jj_1} 
-\Gamma_{(i_2j_2)(i_1j_1)}^{a_2b}(p',p_i,k_i)\,(T^a)_{ba_1}}\nonumber\\
\lefteqn{\mspace{150mu}
-\Gamma_{(i_2j_1)(i_1j_2)}^{a_2a_1}(p',p_i,k_i)\,T^a(\phi)\,\Big\}\ ,}\nonumber
\end{alignat}
\end{subequations}
where $p{=}xP$.
The expression between braces $\{\,\cdots\,\}$ vanishes due to the Ward-Takahashi identity~\eqref{WaTa} for a zero momentum gluon $p^\mu{=}0^\mu$
(this is a reflection of color charge conservation). 
The result shows that the resummation of all single-gluon insertions amounts to multiplying the external legs of the basic Green's function by factors that depend only on the momentum fraction $x$ of the inserted gluon.
The integration over this momentum fraction can be performed independently from the internal structure of the Green's function that does not depend on $x$.

It is interesting to consider the Fourier transform of expression~\eqref{insertions-result}.
Integrating over all momentum fractions $x$ and explicitly writing the
\begin{equation}\label{INT}
\int\frac{\mathrm d\eta^-}{2\pi}\ e^{ixP^+(\eta^--\xi^-)}\,n{\cdot}A^a(\eta)\ ,
\end{equation}
that is contained in the correlator $\Phi^a(p'{-}p,p)$,
equation~\eqref{QQAcorr}, it becomes
\begin{equation}\begin{split}\label{wrd}
\int&\frac{\mathrm d\eta^-}{2\pi}\int\mathrm dx\ e^{ixP^+(\eta^--\xi^-)}
\sum_{\text{insertions}}n{\cdot}A^a(\eta)\ \ P_\mu\,
\Gamma_{(i_2j_2)(i_1j_1)}^{(\mu a)a_2a_1}(p;p'{-}p,p_i,k_i)\\
&=\big\{\,\big(U_{[+\infty;\xi]}\big)_{j_2j}\,
\big(U_{[-\infty;\xi]}\big)_{i_2i}
\big(U_{[+\infty;\xi]}\big)_{a_2b}
\big(U_{[\xi;-\infty]}\big)_{i'i_1}
\big(U_{[\xi;+\infty]}\big)_{j'j_1}
\big(U_{[\xi;-\infty]}\big)_{b'a_1}\,\big\}^g\,
\Gamma_{(ij)(i'j')}^{bb'}(p',p_i,k_i)\ ,
\end{split}\end{equation}
where $\{\,\cdots\,\}^g$ refers to the order $g^1$ term of the combination of Wilson lines.
These Wilson lines are given by
\begin{equation}\label{WILSONline}
U_{[\alpha;\beta]}=\mathcal P\exp
\Big[-ig\int_\alpha^\beta\mathrm d\eta^-\ n{\cdot}A^a(\eta)\ T^a\,\Big]\ ,
\end{equation}
in the fundamental representation for fermions and in the adjoint representation for gluons (cf.~\eqref{infini1} and~\eqref{infini2}).

So far we have taken all collinear interactions of a single gluon into account.
The expression for two-gluon interactions gives the order $g^2$ analogue of~\eqref{wrd} upon Fourier transformation,~etc.
Thus, we have found that resumming \emph{all} collinear interactions amounts to attaching Wilson lines to every external leg other than the one corresponding to $\phi(p')$, 
where the dimension of the representations and the direction of the Wilson lines depend on the nature of the external legs
\begin{equation}
\big(U_{[+\infty;\xi]}\big)_{j_2j}\,
\big(U_{[-\infty;\xi]}\big)_{i_2i}
\big(U_{[+\infty;\xi]}\big)_{a_2b}\,
\Gamma_{(ij)(i'j')}^{bb'}(p',p_i,k_i)\,
\big(U_{[\xi;-\infty]}\big)_{i'i_1}
\big(U_{[\xi;+\infty]}\big)_{j'j_1}
\big(U_{[\xi;-\infty]}\big)_{b'a_1}\ .\label{VLAFLIP}
\end{equation}
Note that this refers only to the \emph{color} structure of the Wilson lines,
since the field operators $A(\eta)$ appear in the correlator $\Phi(p')$ for parton $\phi(p')$.

The procedure outlined above applies to the amplitude on the l.h.s.\ of the cut in the diagram for the scattering cross section. 
The analogous results for the r.h.s.\ of the cut,
which corresponds to the hermitian conjugate of an amplitude,
can be obtained from~\eqref{VLAFLIP} in the obvious way.
That is, by hermitian conjugation.
Since on the r.h.s.\ of the cut $\xi{=}0$ in the exponent in equation~\eqref{INT} and since hermitian conjugation reverses the direction of the Wilson lines we get the results summarized in Table~\ref{TABLE1}.
The emergence of transverse pieces requires extension to higher twist and will proceed along the same lines as discussed in Ref.~\cite{Boer:1999si} or Ref.~\cite{Boer:2003cm}.
                                                                                
We would like to emphasize that the results obtained so far do not depend on the internal structure of the Green's functions.
Only the nature of the external partons of the Green's functions matter.
However, expression~\eqref{VLAFLIP} is not the final answer yet.
This is because the Wilson lines cannot be absorbed directly into the correlator $\Phi(p')$ of the parton $\phi(p')$ as it stands in~\eqref{VLAFLIP}.
To achieve that one should pull the Wilson lines through the hard parts
$\Gamma(p',p_i,k_i)$, 
moving all their color structures to the external leg corresponding to the parton $\phi(p')$.
This can be done by using color flow arguments such as~\eqref{ColorFlow}.
It is at this point that the internal color structure of the Green's function becomes important and that the subprocess-dependence comes in.
Once that all the Wilson lines have been moved to the external leg corresponding to parton $\phi(p')$,
they combine into the full gauge-link.
This gauge-link can now be absorbed in the correlator $\Phi(p')$,
thereby defining a new and gauge-invariant correlator.
This procedure must be followed for each correlator in the scattering process
(for difficulties associated with intertwined insertions from different
correlators we refer to Ref.~\cite{Boer:1999si}).
Absorbing all the Wilson lines in the proper correlators, 
one has obtained the gauge-invariant expression for the process represented by the particular diagram $\Gamma(p',p_i,k_i)$.
In this derivation we have used the axial gauge,
such that there are no ghost fields.
However, in other gauges these should properly be taken into account.
In Ref.~\cite{Pijlman:2006vm} some of the subtleties arising when the Feynman gauge is used are discussed.
In that reference the same results as described here are obtained by explicitly working out several examples in the Feynman gauge.

\section{Conclusions}

High energy hadronic scattering processes can be described as convolutions of
nonperturbative parton distribution and fragmentation functions with a
perturbative hard partonic subprocess. The parton distribution and 
fragmentation functions are nonlocal matrix elements of the parton 
field operators. These matrix elements contain gauge-links, path-ordered 
exponentials, rendering them properly gauge-invariant.
The gauge-links are obtained by resumming all collinear gluon exchanges 
between the hard part and the distribution (fragmentation) function.
In this case collinearity refers to the momentum direction of the hadron 
to which the distribution or fragmentation function belongs.
For many observables the parton momentum along the hadron momentum is the only relevant momentum component, 
the other components being integrated over.
In these collinear distribution and fragmentation functions the field operators 
in the matrix elements are separated along the light-cone.
The gauge-link built from collinear gluons is, then, 
a simple Wilson line along this light-cone direction.

The colored fields appearing in the hadronic matrix elements of TMD distribution and fragmentation functions are not only separated in the light-cone direction,
but also in the transverse direction 
(conjugate to the measured intrinsic transverse momentum of partons). 
In that case there is no unique way of connecting the fields with a gauge-link.
Which gauge-link appears between the parton fields is determined by the resummation of all collinear gluon exchanges and, as such, 
is determined by the hard part of the process.
It turns out that the resummation leads to gauge-links running to infinity in 
the direction conjugate to the, 
in the high-energy limit light-like, hadron momentum. 
Going beyond leading twist, 
contributions are found that close the gauge-links at (plus or minus) infinity. 

In the description of single-spin asymmetries time-reversal behavior is a distinctive feature.
One needs an odd number
(typically one) of $T$-odd distribution or fragmentation functions.
$T$-odd functions appear when the intrinsic transverse momentum of the partons is taken into account.
Since the full transverse momentum dependence requires a treatment beyond leading twist, 
it is more appropriate to look at transverse moments, 
i.e.\ functions weighed with some power of the transverse momentum~\cite{Boer:1997nt,Boer:2003cm}.
These functions appear at subleading order in an expansion in the 
inverse hard scale in integrated cross sections, 
but also at leading order in appropriately weighted cross sections.
In particular for the transverse moments of $T$-odd distribution
functions one uniquely picks up contributions of gluonic pole matrix elements. 
These contain transverse gluon fields at infinity that find their origin in the gauge-links.
The reason is that gluonic pole matrix elements typically have opposite time-reversal behavior
compared to the matrix elements without the zero-momentum gluon fields. 

In SIDIS and DY the resummation of all collinear gluon exchanges leads to the familiar future ($\mathcal U^{[+]}$) and past ($\mathcal U^{[-]}$) pointing Wilson lines, respectively,
leading for the first transverse moments to $T$-odd correlators that differ by a sign.
In these examples the resummation of all collinear gluon exchanges is straightforward,
involving only final state interactions in SIDIS and only initial state interactions in DY.
However, when considering more complicated hadronic processes,
such as $p^\uparrow p{\rightarrow}\pi\pi X$,
the resummation of all collinear gluon exchanges grows increasingly tedious, 
quickly becoming too involved for practical purposes.
In this paper we have presented a prescription for deriving the resulting gauge-links simply by considering the color structure of the hard part and the nature of its external particles.
This prescription leads to a shortcut for the resummation providing a practical tool that makes the calculation of the gauge-links straightforward.
Using this algorithm, the gauge-links appearing in all the tree-level $2{\rightarrow}2$ parton scattering processes through strong interactions were calculated and are enumerated in the appendix.

The procedure shows that it is the flow of color rather than the
flow of fermion number which determines the structure of gauge-links and,
thus, 
also the way in which gluonic pole matrix elements contribute. 
For instance, 
in $eq{\rightarrow}eq$ via photon exchange the color of the quark flows into
the final state resulting in a $\mathcal U^{[+]}$-link.
In contrast, in $\bar q q{\rightarrow}\bar q q$ scattering via one-gluon exchange the dominant contribution is color annihilation in the initial state, 
with $1/N_c$ corrections.
This leads to a gauge-link that is predominantly in the $\mathcal U^{[-]}$ direction.
Another example is $q\bar q{\rightarrow}e\bar e$ via annihilation into a photon. Here there is color-annihilation leading to a $\mathcal U^{[-]}$.
In $q\bar q{\rightarrow}q\bar q$ via annihilation into a gluon,
on the other hand,
the color flow is mostly into the final state (with $1/N_c$ corrections) leading to a gauge-link that is predominantly in the $\mathcal U^{[+]}$ direction
(times a traced loop that does not contribute to transverse moments).

In an earlier paper~\cite{Bacchetta:2005rm} it was shown how the expression for single spin asymmetries can still be cast into a convolution of distributions, fragmentations and hard parts even when the gauge-links are incorporated.
The hard parts are referred to as gluonic pole cross sections. 
Using the results obtained here we will be able to present all possible gluonic pole cross sections appearing in processes like proton-proton scattering in a forthcoming publication.
The use of these gluonic pole cross sections is important in processes like 
2-jet or inclusive 2-pion production in proton-proton scattering,
while it may also modify the standard approach used for single-pion production~\cite{Anselmino:1994tv,Anselmino:1998yz}.

\begin{acknowledgments}
We acknowledge discussions with D.~Boer. 
Part of this work was supported by the foundation for Fundamental 
Research of Matter (FOM) and the National Organization for Scientific 
Research (NWO).
\end{acknowledgments}

\appendix

\section{Gauge links for $2\rightarrow 2$ parton processes\label{PPscat}}

In Table~\ref{Tqq2qq}-\ref{Tgg2gg} we enumerate all the gauge-links that 
appear in hard scattering processes with $2{\rightarrow}2$ partonic 
subprocesses at tree-level. 
For the fragmentation correlators we use the symbolic 
notation discussed in Example~B in section~\ref{VOORBEELDEN}.
In the case of gluon (fragmentation) correlators the gauge-links might 
depend on whether the gluons couple to a quark or an antiquark. 
If both situations occur in the same diagram, we will distinguish these 
by writing $\Phi_{g(q)}$ ($\Delta_{g(q)}$) and 
$\Phi_{g(\bar q)}$ ($\Delta_{g(\bar q)}$).
The diagrams involving 4-gluon vertices can be obtained from the diagrams 
in Table~\ref{Tgg2gg} in the way discussed in Example~E in 
section~\ref{VOORBEELDEN}.

\bibliographystyle{apsrev}
\bibliography{references}

\begin{thebibliography}{42}
\expandafter\ifx\csname natexlab\endcsname\relax\def\natexlab#1{#1}\fi
\expandafter\ifx\csname bibnamefont\endcsname\relax
  \def\bibnamefont#1{#1}\fi
\expandafter\ifx\csname bibfnamefont\endcsname\relax
  \def\bibfnamefont#1{#1}\fi
\expandafter\ifx\csname citenamefont\endcsname\relax
  \def\citenamefont#1{#1}\fi
\expandafter\ifx\csname url\endcsname\relax
  \def\url#1{\texttt{#1}}\fi
\expandafter\ifx\csname urlprefix\endcsname\relax\def\urlprefix{URL }\fi
\providecommand{\bibinfo}[2]{#2}
\providecommand{\eprint}[2][]{\url{#2}}

\bibitem[{\citenamefont{Adams et~al.}(1991{\natexlab{a}})}]{Adams:1991rw}
\bibinfo{author}{\bibfnamefont{D.~L.} \bibnamefont{Adams}} \bibnamefont{et~al.}
  (\bibinfo{collaboration}{E581}), \bibinfo{journal}{Phys. Lett.}
  \textbf{\bibinfo{volume}{B261}}, \bibinfo{pages}{201}
  (\bibinfo{year}{1991}{\natexlab{a}}).

\bibitem[{\citenamefont{Adams et~al.}(1991{\natexlab{b}})}]{Adams:1991cs}
\bibinfo{author}{\bibfnamefont{D.~L.} \bibnamefont{Adams}} \bibnamefont{et~al.}
  (\bibinfo{collaboration}{FNAL-E704}), \bibinfo{journal}{Phys. Lett.}
  \textbf{\bibinfo{volume}{B264}}, \bibinfo{pages}{462}
  (\bibinfo{year}{1991}{\natexlab{b}}).

\bibitem[{\citenamefont{Bravar et~al.}(1996)}]{Bravar:1996ki}
\bibinfo{author}{\bibfnamefont{A.}~\bibnamefont{Bravar}} \bibnamefont{et~al.}
  (\bibinfo{collaboration}{Fermilab E704}), \bibinfo{journal}{Phys. Rev. Lett.}
  \textbf{\bibinfo{volume}{77}}, \bibinfo{pages}{2626} (\bibinfo{year}{1996}).

\bibitem[{\citenamefont{Airapetian et~al.}(2001)}]{Airapetian:2001eg}
\bibinfo{author}{\bibfnamefont{A.}~\bibnamefont{Airapetian}}
  \bibnamefont{et~al.} (\bibinfo{collaboration}{HERMES}),
  \bibinfo{journal}{Phys. Rev.} \textbf{\bibinfo{volume}{D64}},
  \bibinfo{pages}{097101} (\bibinfo{year}{2001}), \eprint{hep-ex/0104005}.

\bibitem[{\citenamefont{Adler et~al.}(2003)}]{Adler:2003pb}
\bibinfo{author}{\bibfnamefont{S.~S.} \bibnamefont{Adler}} \bibnamefont{et~al.}
  (\bibinfo{collaboration}{PHENIX}), \bibinfo{journal}{Phys. Rev. Lett.}
  \textbf{\bibinfo{volume}{91}}, \bibinfo{pages}{241803}
  (\bibinfo{year}{2003}), \eprint{hep-ex/0304038}.

\bibitem[{\citenamefont{Adams et~al.}(2004)}]{Adams:2003fx}
\bibinfo{author}{\bibfnamefont{J.}~\bibnamefont{Adams}} \bibnamefont{et~al.}
  (\bibinfo{collaboration}{STAR}), \bibinfo{journal}{Phys. Rev. Lett.}
  \textbf{\bibinfo{volume}{92}}, \bibinfo{pages}{171801}
  (\bibinfo{year}{2004}), \eprint{hep-ex/0310058}.

\bibitem[{\citenamefont{Airapetian et~al.}(2005)}]{Airapetian:2004tw}
\bibinfo{author}{\bibfnamefont{A.}~\bibnamefont{Airapetian}}
  \bibnamefont{et~al.} (\bibinfo{collaboration}{HERMES}),
  \bibinfo{journal}{Phys. Rev. Lett.} \textbf{\bibinfo{volume}{94}},
  \bibinfo{pages}{012002} (\bibinfo{year}{2005}), \eprint{hep-ex/0408013}.

\bibitem[{\citenamefont{Hagiwara et~al.}(1983)\citenamefont{Hagiwara, Hikasa,
  and Kai}}]{Hagiwara:1982cq}
\bibinfo{author}{\bibfnamefont{K.}~\bibnamefont{Hagiwara}},
  \bibinfo{author}{\bibfnamefont{K.-i.} \bibnamefont{Hikasa}},
  \bibnamefont{and} \bibinfo{author}{\bibfnamefont{N.}~\bibnamefont{Kai}},
  \bibinfo{journal}{Phys. Rev.} \textbf{\bibinfo{volume}{D27}},
  \bibinfo{pages}{84} (\bibinfo{year}{1983}).

\bibitem[{\citenamefont{Sivers}(1990)}]{Sivers:1989cc}
\bibinfo{author}{\bibfnamefont{D.~W.} \bibnamefont{Sivers}},
  \bibinfo{journal}{Phys. Rev.} \textbf{\bibinfo{volume}{D41}},
  \bibinfo{pages}{83} (\bibinfo{year}{1990}).

\bibitem[{\citenamefont{Sivers}(1991)}]{Sivers:1990fh}
\bibinfo{author}{\bibfnamefont{D.~W.} \bibnamefont{Sivers}},
  \bibinfo{journal}{Phys. Rev.} \textbf{\bibinfo{volume}{D43}},
  \bibinfo{pages}{261} (\bibinfo{year}{1991}).

\bibitem[{\citenamefont{Qiu and Sterman}(1991)}]{Qiu:1991pp}
\bibinfo{author}{\bibfnamefont{J.-w.} \bibnamefont{Qiu}} \bibnamefont{and}
  \bibinfo{author}{\bibfnamefont{G.}~\bibnamefont{Sterman}},
  \bibinfo{journal}{Phys. Rev. Lett.} \textbf{\bibinfo{volume}{67}},
  \bibinfo{pages}{2264} (\bibinfo{year}{1991}).

\bibitem[{\citenamefont{Qiu and Sterman}(1992)}]{Qiu:1991wg}
\bibinfo{author}{\bibfnamefont{J.-w.} \bibnamefont{Qiu}} \bibnamefont{and}
  \bibinfo{author}{\bibfnamefont{G.}~\bibnamefont{Sterman}},
  \bibinfo{journal}{Nucl. Phys.} \textbf{\bibinfo{volume}{B378}},
  \bibinfo{pages}{52} (\bibinfo{year}{1992}).

\bibitem[{\citenamefont{Collins}(1993)}]{Collins:1992kk}
\bibinfo{author}{\bibfnamefont{J.~C.} \bibnamefont{Collins}},
  \bibinfo{journal}{Nucl. Phys.} \textbf{\bibinfo{volume}{B396}},
  \bibinfo{pages}{161} (\bibinfo{year}{1993}), \eprint{hep-ph/9208213}.

\bibitem[{\citenamefont{Kanazawa and Koike}(2000)}]{Kanazawa:2000hz}
\bibinfo{author}{\bibfnamefont{Y.}~\bibnamefont{Kanazawa}} \bibnamefont{and}
  \bibinfo{author}{\bibfnamefont{Y.}~\bibnamefont{Koike}},
  \bibinfo{journal}{Phys. Lett.} \textbf{\bibinfo{volume}{B478}},
  \bibinfo{pages}{121} (\bibinfo{year}{2000}), \eprint{hep-ph/0001021}.

\bibitem[{\citenamefont{Efremov and Teryaev}(1984)}]{Efremov}
\bibinfo{author}{\bibfnamefont{A.~V.} \bibnamefont{Efremov}} \bibnamefont{and}
  \bibinfo{author}{\bibfnamefont{O.~V.} \bibnamefont{Teryaev}},
  \bibinfo{journal}{Sov. J. Nucl. Phys.} \textbf{\bibinfo{volume}{39}},
  \bibinfo{pages}{962} (\bibinfo{year}{1984}).

\bibitem[{\citenamefont{Ji}(1992)}]{Ji:1992eu}
\bibinfo{author}{\bibfnamefont{X.-D.} \bibnamefont{Ji}},
  \bibinfo{journal}{Phys. Lett.} \textbf{\bibinfo{volume}{B289}},
  \bibinfo{pages}{137} (\bibinfo{year}{1992}).

\bibitem[{\citenamefont{Brodsky
  et~al.}(2002{\natexlab{a}})\citenamefont{Brodsky, Hwang, and
  Schmidt}}]{Brodsky:2002cx}
\bibinfo{author}{\bibfnamefont{S.~J.} \bibnamefont{Brodsky}},
  \bibinfo{author}{\bibfnamefont{D.~S.} \bibnamefont{Hwang}}, \bibnamefont{and}
  \bibinfo{author}{\bibfnamefont{I.}~\bibnamefont{Schmidt}},
  \bibinfo{journal}{Phys. Lett.} \textbf{\bibinfo{volume}{B530}},
  \bibinfo{pages}{99} (\bibinfo{year}{2002}{\natexlab{a}}),
  \eprint{hep-ph/0201296}.

\bibitem[{\citenamefont{Collins}(2002)}]{Collins:2002kn}
\bibinfo{author}{\bibfnamefont{J.~C.} \bibnamefont{Collins}},
  \bibinfo{journal}{Phys. Lett.} \textbf{\bibinfo{volume}{B536}},
  \bibinfo{pages}{43} (\bibinfo{year}{2002}), \eprint{hep-ph/0204004}.

\bibitem[{\citenamefont{Brodsky
  et~al.}(2002{\natexlab{b}})\citenamefont{Brodsky, Hwang, and
  Schmidt}}]{Brodsky:2002rv}
\bibinfo{author}{\bibfnamefont{S.~J.} \bibnamefont{Brodsky}},
  \bibinfo{author}{\bibfnamefont{D.~S.} \bibnamefont{Hwang}}, \bibnamefont{and}
  \bibinfo{author}{\bibfnamefont{I.}~\bibnamefont{Schmidt}},
  \bibinfo{journal}{Nucl. Phys.} \textbf{\bibinfo{volume}{B642}},
  \bibinfo{pages}{344} (\bibinfo{year}{2002}{\natexlab{b}}),
  \eprint{hep-ph/0206259}.

\bibitem[{\citenamefont{Belitsky et~al.}(2003)\citenamefont{Belitsky, Ji, and
  Yuan}}]{Belitsky:2002sm}
\bibinfo{author}{\bibfnamefont{A.~V.} \bibnamefont{Belitsky}},
  \bibinfo{author}{\bibfnamefont{X.}~\bibnamefont{Ji}}, \bibnamefont{and}
  \bibinfo{author}{\bibfnamefont{F.}~\bibnamefont{Yuan}},
  \bibinfo{journal}{Nucl. Phys.} \textbf{\bibinfo{volume}{B656}},
  \bibinfo{pages}{165} (\bibinfo{year}{2003}), \eprint{hep-ph/0208038}.

\bibitem[{\citenamefont{Boer et~al.}(2003)\citenamefont{Boer, Mulders, and
  Pijlman}}]{Boer:2003cm}
\bibinfo{author}{\bibfnamefont{D.}~\bibnamefont{Boer}},
  \bibinfo{author}{\bibfnamefont{P.~J.} \bibnamefont{Mulders}},
  \bibnamefont{and} \bibinfo{author}{\bibfnamefont{F.}~\bibnamefont{Pijlman}},
  \bibinfo{journal}{Nucl. Phys.} \textbf{\bibinfo{volume}{B667}},
  \bibinfo{pages}{201} (\bibinfo{year}{2003}), \eprint{hep-ph/0303034}.

\bibitem[{\citenamefont{Bomhof et~al.}(2004)\citenamefont{Bomhof, Mulders, and
  Pijlman}}]{Bomhof:2004aw}
\bibinfo{author}{\bibfnamefont{C.~J.} \bibnamefont{Bomhof}},
  \bibinfo{author}{\bibfnamefont{P.~J.} \bibnamefont{Mulders}},
  \bibnamefont{and} \bibinfo{author}{\bibfnamefont{F.}~\bibnamefont{Pijlman}},
  \bibinfo{journal}{Phys. Lett.} \textbf{\bibinfo{volume}{B596}},
  \bibinfo{pages}{277} (\bibinfo{year}{2004}), \eprint{hep-ph/0406099}.

\bibitem[{\citenamefont{Bacchetta et~al.}(2005)\citenamefont{Bacchetta, Bomhof,
  Mulders, and Pijlman}}]{Bacchetta:2005rm}
\bibinfo{author}{\bibfnamefont{A.}~\bibnamefont{Bacchetta}},
  \bibinfo{author}{\bibfnamefont{C.~J.} \bibnamefont{Bomhof}},
  \bibinfo{author}{\bibfnamefont{P.~J.} \bibnamefont{Mulders}},
  \bibnamefont{and} \bibinfo{author}{\bibfnamefont{F.}~\bibnamefont{Pijlman}}
  (\bibinfo{year}{2005}), \eprint{hep-ph/0505268}.

\bibitem[{\citenamefont{Pijlman}(2006)}]{Pijlman:2006vm}
\bibinfo{author}{\bibfnamefont{F.}~\bibnamefont{Pijlman}}, Ph.D. thesis,
  \bibinfo{school}{Vrije Universiteit Amsterdam} (\bibinfo{year}{2006}).

\bibitem[{\citenamefont{Soper}(1977)}]{Soper:1976jc}
\bibinfo{author}{\bibfnamefont{D.~E.} \bibnamefont{Soper}},
  \bibinfo{journal}{Phys. Rev.} \textbf{\bibinfo{volume}{D15}},
  \bibinfo{pages}{1141} (\bibinfo{year}{1977}).

\bibitem[{\citenamefont{Soper}(1979)}]{Soper:1979fq}
\bibinfo{author}{\bibfnamefont{D.~E.} \bibnamefont{Soper}},
  \bibinfo{journal}{Phys. Rev. Lett.} \textbf{\bibinfo{volume}{43}},
  \bibinfo{pages}{1847} (\bibinfo{year}{1979}).

\bibitem[{\citenamefont{Ralston and Soper}(1979)}]{Ralston:1979ys}
\bibinfo{author}{\bibfnamefont{J.~P.} \bibnamefont{Ralston}} \bibnamefont{and}
  \bibinfo{author}{\bibfnamefont{D.~E.} \bibnamefont{Soper}},
  \bibinfo{journal}{Nucl. Phys.} \textbf{\bibinfo{volume}{B152}},
  \bibinfo{pages}{109} (\bibinfo{year}{1979}).

\bibitem[{\citenamefont{Collins and Soper}(1982)}]{Collins:1981uw}
\bibinfo{author}{\bibfnamefont{J.~C.} \bibnamefont{Collins}} \bibnamefont{and}
  \bibinfo{author}{\bibfnamefont{D.~E.} \bibnamefont{Soper}},
  \bibinfo{journal}{Nucl. Phys.} \textbf{\bibinfo{volume}{B194}},
  \bibinfo{pages}{445} (\bibinfo{year}{1982}).

\bibitem[{\citenamefont{Collins et~al.}(1982)\citenamefont{Collins, Soper, and
  Sterman}}]{Collins:1981tt}
\bibinfo{author}{\bibfnamefont{J.~C.} \bibnamefont{Collins}},
  \bibinfo{author}{\bibfnamefont{D.~E.} \bibnamefont{Soper}}, \bibnamefont{and}
  \bibinfo{author}{\bibfnamefont{G.}~\bibnamefont{Sterman}},
  \bibinfo{journal}{Phys. Lett.} \textbf{\bibinfo{volume}{B109}},
  \bibinfo{pages}{388} (\bibinfo{year}{1982}).

\bibitem[{\citenamefont{Collins et~al.}(1983)\citenamefont{Collins, Soper, and
  Sterman}}]{Collins:1982wa}
\bibinfo{author}{\bibfnamefont{J.~C.} \bibnamefont{Collins}},
  \bibinfo{author}{\bibfnamefont{D.~E.} \bibnamefont{Soper}}, \bibnamefont{and}
  \bibinfo{author}{\bibfnamefont{G.}~\bibnamefont{Sterman}},
  \bibinfo{journal}{Nucl. Phys.} \textbf{\bibinfo{volume}{B223}},
  \bibinfo{pages}{381} (\bibinfo{year}{1983}).

\bibitem[{\citenamefont{Jaffe and Ji}(1992)}]{Jaffe:1991ra}
\bibinfo{author}{\bibfnamefont{R.~L.} \bibnamefont{Jaffe}} \bibnamefont{and}
  \bibinfo{author}{\bibfnamefont{X.-D.} \bibnamefont{Ji}},
  \bibinfo{journal}{Nucl. Phys.} \textbf{\bibinfo{volume}{B375}},
  \bibinfo{pages}{527} (\bibinfo{year}{1992}).

\bibitem[{\citenamefont{Mulders and Tangerman}(1996)}]{Mulders:1995dh}
\bibinfo{author}{\bibfnamefont{P.~J.} \bibnamefont{Mulders}} \bibnamefont{and}
  \bibinfo{author}{\bibfnamefont{R.~D.} \bibnamefont{Tangerman}},
  \bibinfo{journal}{Nucl. Phys.} \textbf{\bibinfo{volume}{B461}},
  \bibinfo{pages}{197} (\bibinfo{year}{1996}), \eprint{hep-ph/9510301}.

\bibitem[{\citenamefont{Efremov and Radyushkin}(1981)}]{Efremov:1978xm}
\bibinfo{author}{\bibfnamefont{A.~V.} \bibnamefont{Efremov}} \bibnamefont{and}
  \bibinfo{author}{\bibfnamefont{A.~V.} \bibnamefont{Radyushkin}},
  \bibinfo{journal}{Theor. Math. Phys.} \textbf{\bibinfo{volume}{44}},
  \bibinfo{pages}{774} (\bibinfo{year}{1981}).

\bibitem[{\citenamefont{Boer and Mulders}(2000)}]{Boer:1999si}
\bibinfo{author}{\bibfnamefont{D.}~\bibnamefont{Boer}} \bibnamefont{and}
  \bibinfo{author}{\bibfnamefont{P.~J.} \bibnamefont{Mulders}},
  \bibinfo{journal}{Nucl. Phys.} \textbf{\bibinfo{volume}{B569}},
  \bibinfo{pages}{505} (\bibinfo{year}{2000}), \eprint{hep-ph/9906223}.

\bibitem[{\citenamefont{Collins et~al.}(1988)\citenamefont{Collins, Soper, and
  Sterman}}]{Collins:1989gx}
\bibinfo{author}{\bibfnamefont{J.~C.} \bibnamefont{Collins}},
  \bibinfo{author}{\bibfnamefont{D.~E.} \bibnamefont{Soper}}, \bibnamefont{and}
  \bibinfo{author}{\bibfnamefont{G.}~\bibnamefont{Sterman}},
  \bibinfo{journal}{Adv. Ser. Direct. High Energy Phys.}
  \textbf{\bibinfo{volume}{5}}, \bibinfo{pages}{1} (\bibinfo{year}{1988}),
  \eprint{hep-ph/0409313}.

\bibitem[{\citenamefont{Jaffe}(1996)}]{Jaffe:1995an}
\bibinfo{author}{\bibfnamefont{R.~L.} \bibnamefont{Jaffe}},
  \bibinfo{journal}{Phys. Lett.} \textbf{\bibinfo{volume}{B365}},
  \bibinfo{pages}{359} (\bibinfo{year}{1996}), \eprint{hep-ph/9509279}.

\bibitem[{\citenamefont{Bashinsky and Jaffe}(1998)}]{Bashinsky:1998if}
\bibinfo{author}{\bibfnamefont{S.~V.} \bibnamefont{Bashinsky}}
  \bibnamefont{and} \bibinfo{author}{\bibfnamefont{R.~L.} \bibnamefont{Jaffe}},
  \bibinfo{journal}{Nucl. Phys.} \textbf{\bibinfo{volume}{B536}},
  \bibinfo{pages}{303} (\bibinfo{year}{1998}), \eprint{hep-ph/9804397}.

\bibitem[{\citenamefont{Mulders and Rodrigues}(2001)}]{Mulders:2000sh}
\bibinfo{author}{\bibfnamefont{P.~J.} \bibnamefont{Mulders}} \bibnamefont{and}
  \bibinfo{author}{\bibfnamefont{J.}~\bibnamefont{Rodrigues}},
  \bibinfo{journal}{Phys. Rev.} \textbf{\bibinfo{volume}{D63}},
  \bibinfo{pages}{094021} (\bibinfo{year}{2001}), \eprint{hep-ph/0009343}.

\bibitem[{\citenamefont{Ji et~al.}(2005)\citenamefont{Ji, Ma, and
  Yuan}}]{Ji:2005nu}
\bibinfo{author}{\bibfnamefont{X.-d.} \bibnamefont{Ji}},
  \bibinfo{author}{\bibfnamefont{J.-P.} \bibnamefont{Ma}}, \bibnamefont{and}
  \bibinfo{author}{\bibfnamefont{F.}~\bibnamefont{Yuan}},
  \bibinfo{journal}{JHEP} \textbf{\bibinfo{volume}{07}}, \bibinfo{pages}{020}
  (\bibinfo{year}{2005}), \eprint{hep-ph/0503015}.

\bibitem[{\citenamefont{Boer and Mulders}(1998)}]{Boer:1997nt}
\bibinfo{author}{\bibfnamefont{D.}~\bibnamefont{Boer}} \bibnamefont{and}
  \bibinfo{author}{\bibfnamefont{P.~J.} \bibnamefont{Mulders}},
  \bibinfo{journal}{Phys. Rev.} \textbf{\bibinfo{volume}{D57}},
  \bibinfo{pages}{5780} (\bibinfo{year}{1998}), \eprint{hep-ph/9711485}.

\bibitem[{\citenamefont{Anselmino et~al.}(1995)\citenamefont{Anselmino,
  Boglione, and Murgia}}]{Anselmino:1994tv}
\bibinfo{author}{\bibfnamefont{M.}~\bibnamefont{Anselmino}},
  \bibinfo{author}{\bibfnamefont{M.}~\bibnamefont{Boglione}}, \bibnamefont{and}
  \bibinfo{author}{\bibfnamefont{F.}~\bibnamefont{Murgia}},
  \bibinfo{journal}{Phys. Lett.} \textbf{\bibinfo{volume}{B362}},
  \bibinfo{pages}{164} (\bibinfo{year}{1995}), \eprint{hep-ph/9503290}.

\bibitem[{\citenamefont{Anselmino and Murgia}(1998)}]{Anselmino:1998yz}
\bibinfo{author}{\bibfnamefont{M.}~\bibnamefont{Anselmino}} \bibnamefont{and}
  \bibinfo{author}{\bibfnamefont{F.}~\bibnamefont{Murgia}},
  \bibinfo{journal}{Phys. Lett.} \textbf{\bibinfo{volume}{B442}},
  \bibinfo{pages}{470} (\bibinfo{year}{1998}), \eprint{hep-ph/9808426}.

\end{thebibliography}

\pagebreak

\begin{table}[h]
\centering
\includegraphics{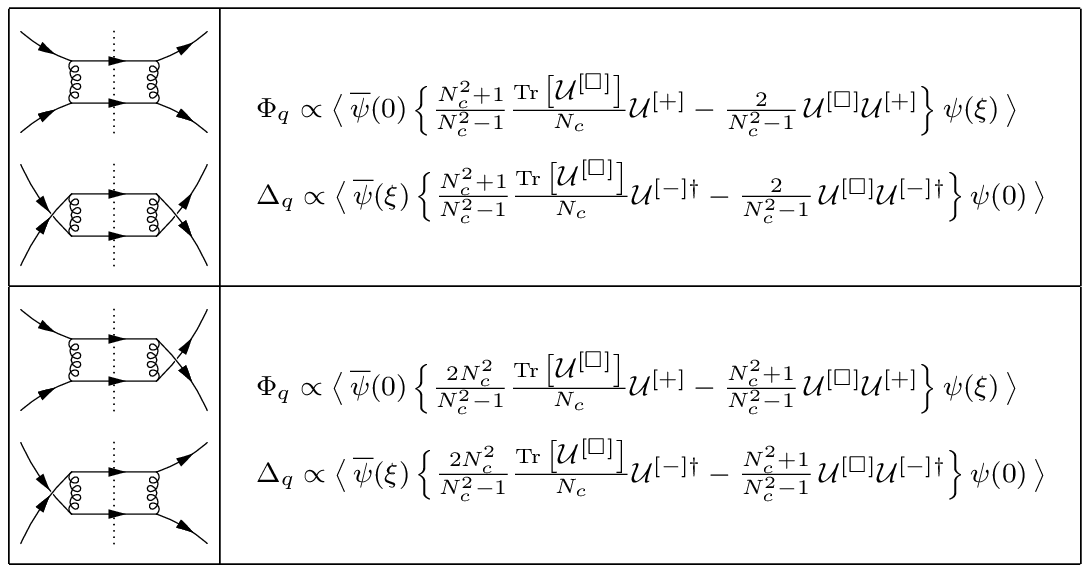}
\caption{Gauge-links appearing in $qq{\rightarrow}qq$.\label{Tqq2qq}}
\end{table}

\begin{table}[h]
\centering
\includegraphics{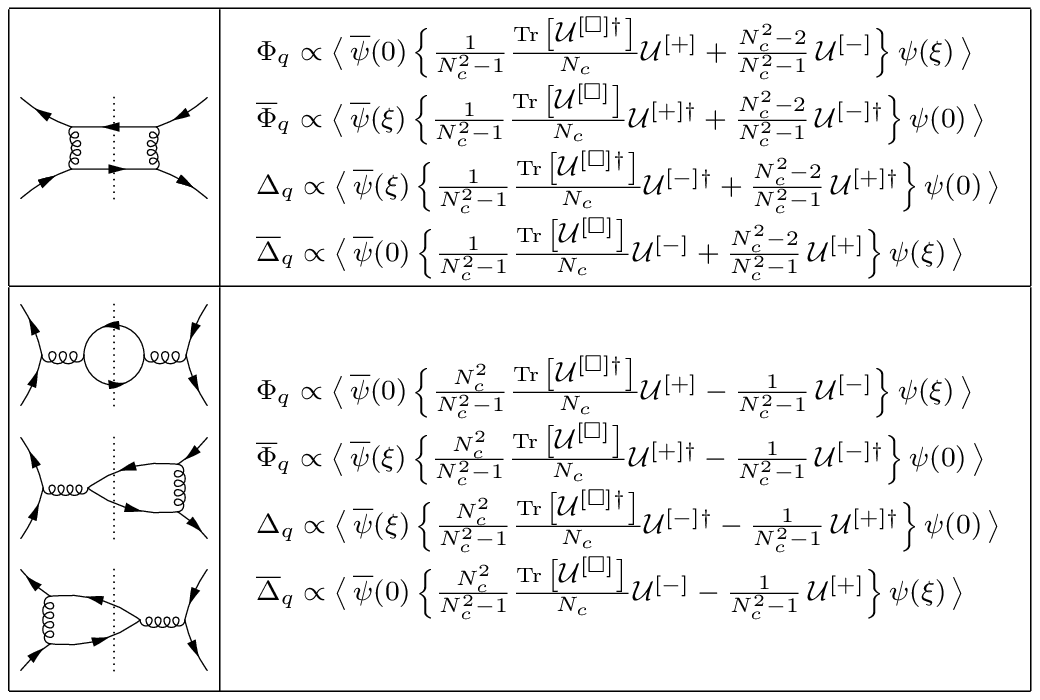}
\caption{Gauge-links appearing in 
$q\bar q{\rightarrow}q\bar q$.\label{Tqqbar2qqbar}}
\end{table}

\begin{table}[h]
\centering
\includegraphics{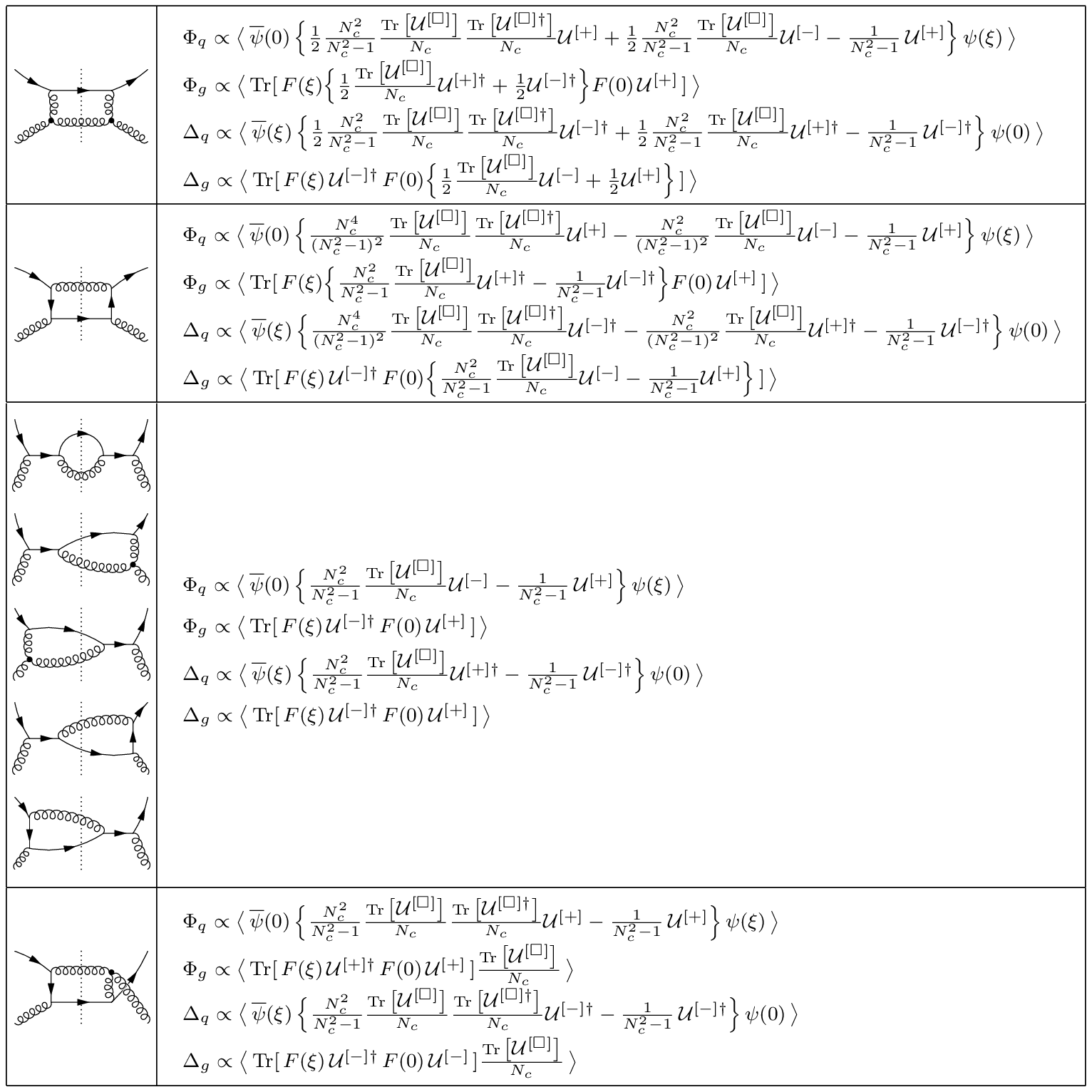}
\caption{Gauge-links appearing in $qg{\rightarrow}qg$.\label{Tqg2qg}}
\end{table}

\begin{table}[h]
\centering
\includegraphics{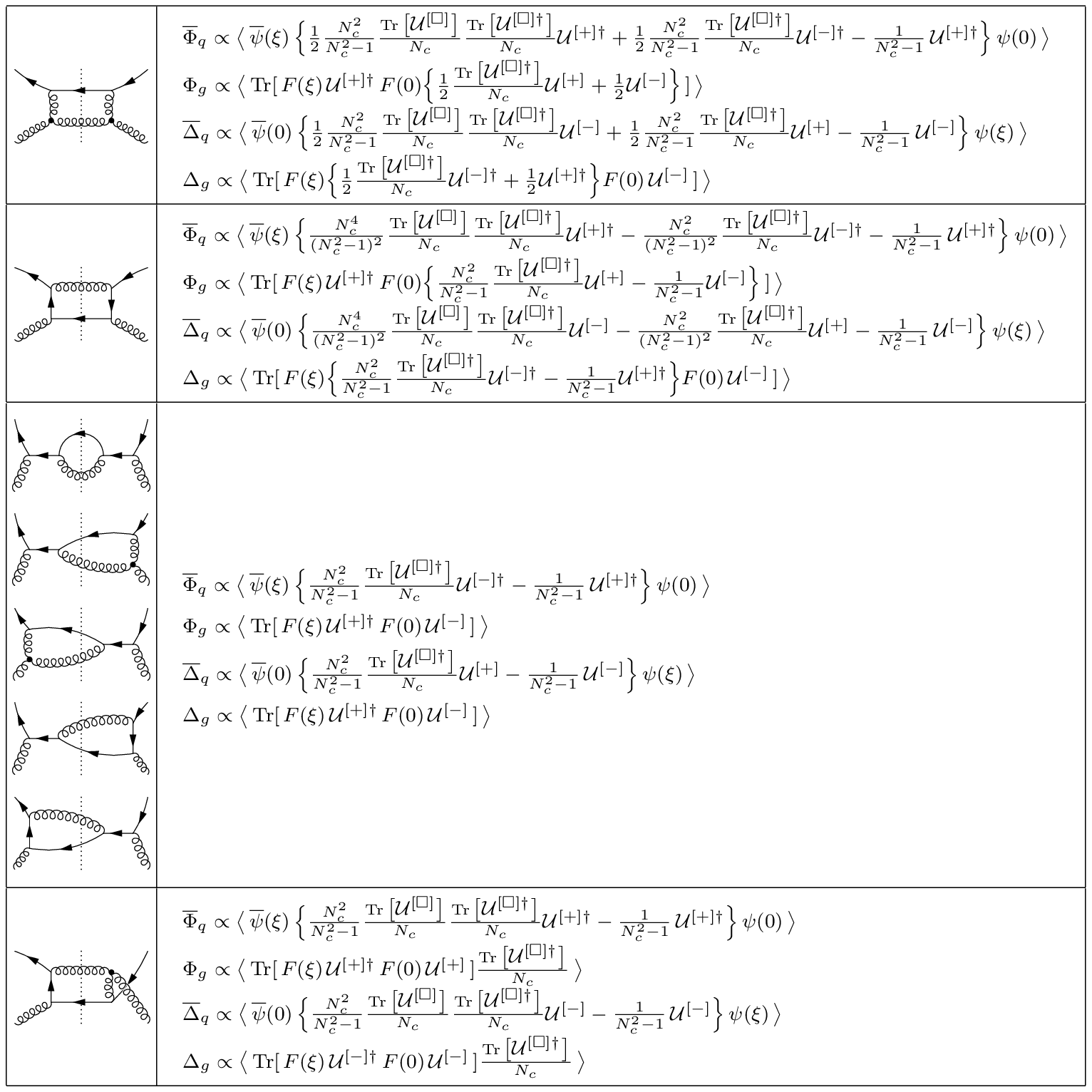}
\caption{Gauge-links appearing in $\bar qg{\rightarrow}\bar qg$.}
\end{table}

\begin{table}[h]
\centering
\includegraphics{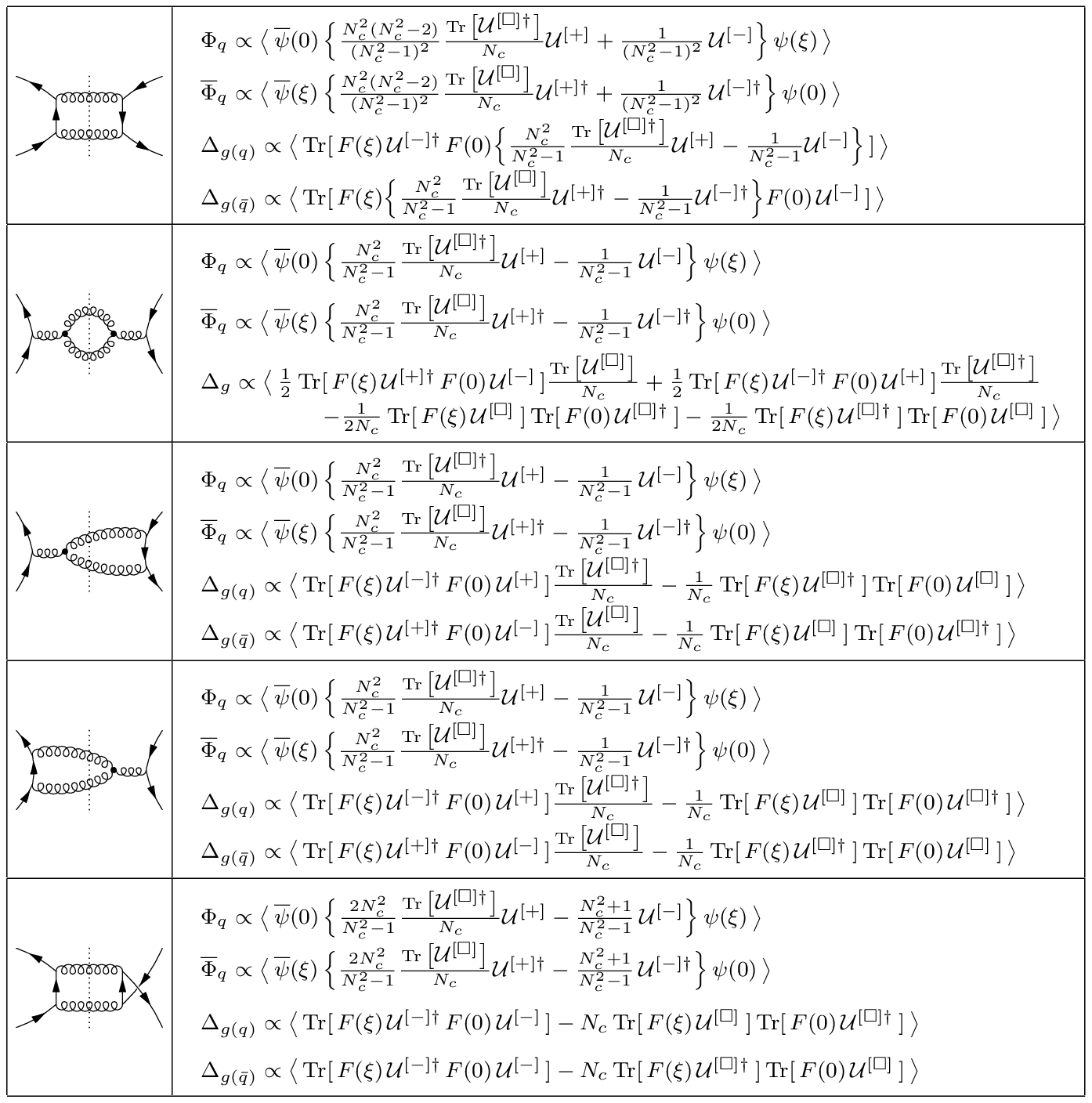}
\caption{Gauge-links appearing in $q\bar q{\rightarrow}gg$.\label{Tqqbar2gg}}
\end{table}

\begin{table}[h]
\centering
\includegraphics{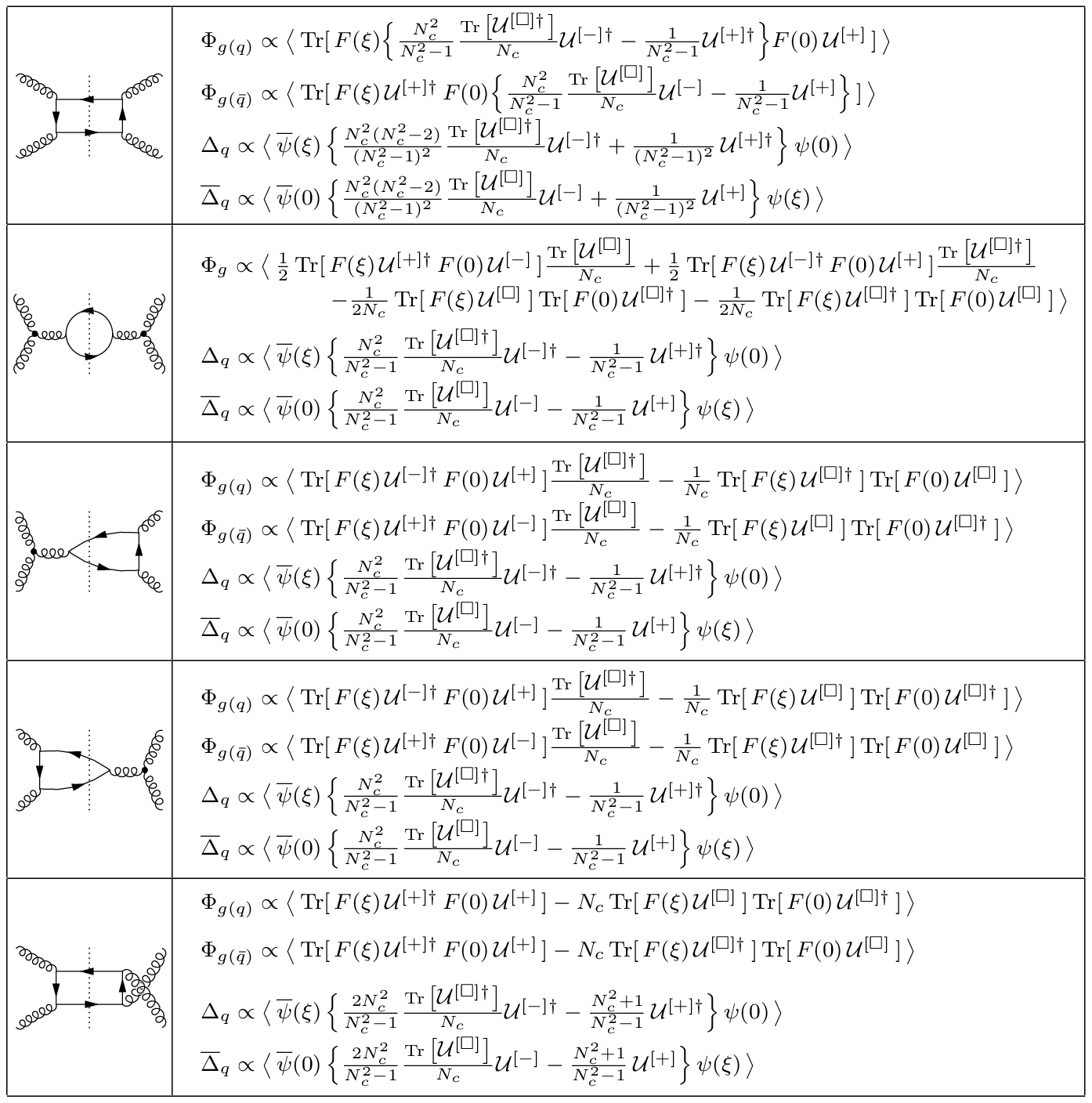}
\caption{Gauge-links appearing in $gg{\rightarrow}q\bar q$.\label{Tgg2qqbar}}
\end{table}

\begin{table}[h]
\centering
\includegraphics{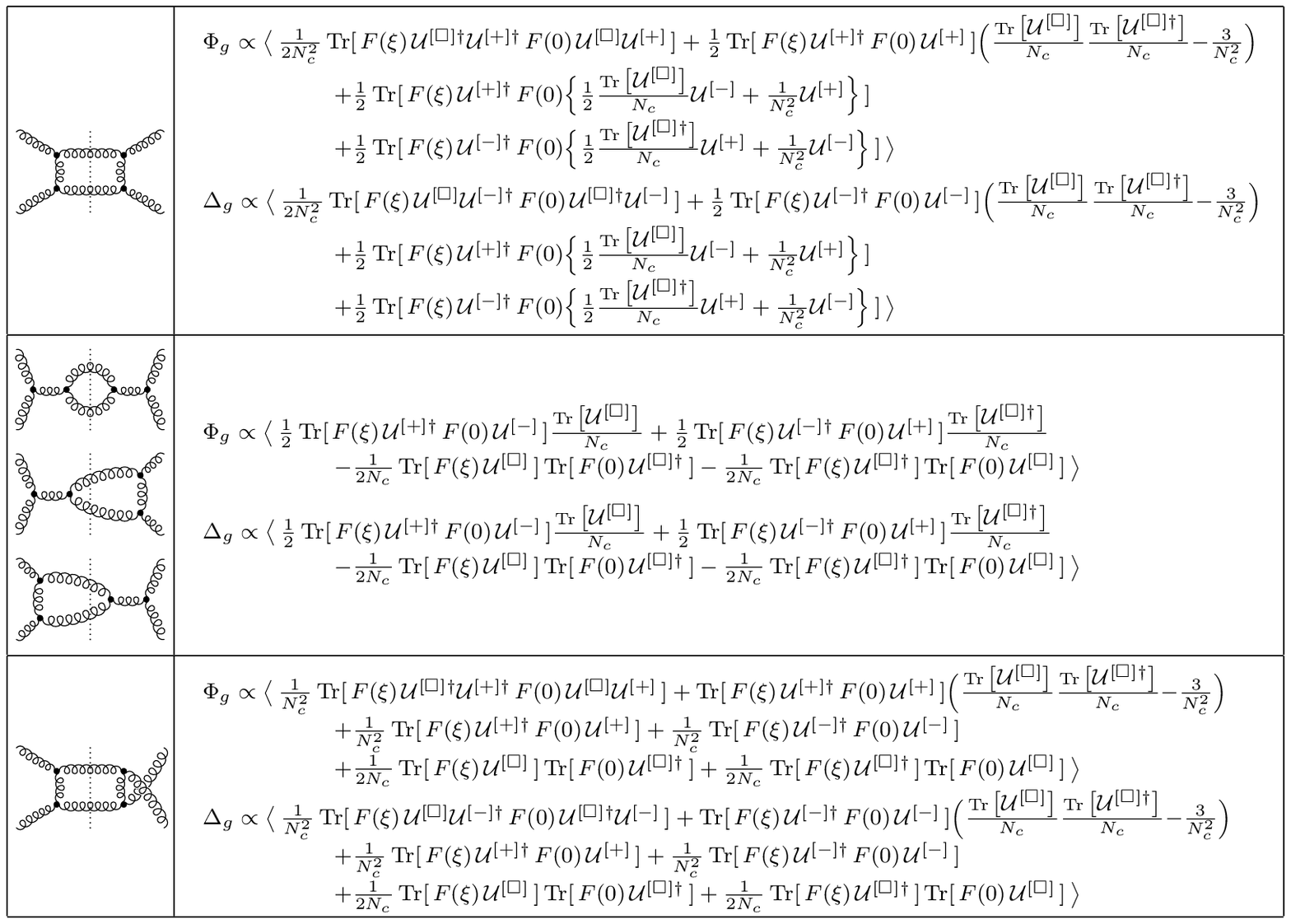}
\caption{Gauge-links appearing in the diagrams without 4-point vertices that contribute in $gg{\rightarrow}gg$.\label{Tgg2gg}}
\end{table}

\end{document}